\documentclass[11pt]{article}
\pdfoutput=1
\usepackage[utf8x]{inputenc}
\usepackage{amssymb,amsmath,mathrsfs,enumerate}
\usepackage{graphicx,rotate,multicol,braket, multirow}
\usepackage{cite}
\usepackage{braket}
\usepackage[normalem]{ulem}
\usepackage{wrapfig}
\usepackage[textfont=it,labelfont=bf]{caption}
\usepackage{bm}
\usepackage[colorlinks=true,
linkcolor=red,
urlcolor=blue,
citecolor=blue]{hyperref}
\usepackage{lineno}
\usepackage{color}
\usepackage{bbold}
\long\def\rpl#1!!#2!!{\textcolor{red}{#1} \textcolor{blue}{#2}}
\def\bar{\overline}
\def\tilde{\widetilde}

\newcommand{\lag}[1]{\mathscr{L}_{\rm #1}}

\evensidemargin=\oddsidemargin
\def\Eqn#1{Eq.\ (\ref{#1})}

\def\Sect#1{Sec.\,\ref{#1}}
\allowdisplaybreaks
\title{\Large\bf 
Democratic three Higgs-doublet models: \\the custodial limit and wrong-sign Yukawa
}

\author{
  \sf 
  Dipankar Das$^{a,}$\footnote{d.das@iiti.ac.in},
  Miguel Levy$^{b,}$\footnote{miguelplevy@tecnico.ulisboa.pt},
  Palash B. Pal$^{c,}$\footnote{palashbaran.pal@saha.ac.in},\\
  \sf
  Anugrah M. Prasad$^{a,}$\footnote{anugrahmprasad@gmail.com},
  Ipsita  Saha$^{d,}$\footnote{ipsita@allduniv.ac.in},
  Ayushi Srivastava$^{a,}$\footnote{srivastavaayushi860@gmail.com}
\\[3mm]
    \small\em
    $^a$Indian Institute of Technology (Indore), Khandwa Road, Simrol,
  Indore 453 552, India \\
  \small\em
  $^b$ Centro de F\'isica Te\'orica de Part\'iculas-CFTP and Departamento de
  F\'isica,  Instituto Superior T\'ecnico,\\  \small\em
  Universidade de Lisboa, Av
  Rovisco Pais, 1, P-1049-001 Lisboa, Portugal \\ 
 \small\em
  $^c$Department of Physics, University of Calcutta, 92 A. P. C. Road,
  Calcutta 700009, India\\
 \small\em
  $^d$Department of Physics, Faculty of Science, University of Allahabad, Old Katra,
  Prayagraj 211002, India
  }

\date{}

\begin{document}

%\pagewiselinenumbers

\maketitle

\renewcommand*{\thefootnote}{\arabic{footnote}}
\setcounter{footnote}{0} 
\begin{abstract}
We study two novel aspects of democratic 3HDMs -- the custodial limit and the possibility of wrong-sign Yukawa couplings.  
In the custodial limit, the democratic 3HDMs can easily negotiate the constraints from the electroweak $T$-parameter.  
We also uncover the possibility of having wrong-sign Yukawa couplings in democratic 3HDMs, as in the case of 2HDMs.  
We show that a democratic 3HDM encompasses all the wrong-sign possibilities entertained by 2HDMs, and has considerably more leeway in the wrong-sign limit as compared to the 2HDM case.  
Our study underscores the importance of reporting analysis in the kappa-formalism without any implicit assumptions on the signs of the kappas.
\end{abstract}

\bigskip
%===============================================

\section{Introduction} \label{s:intro}
%%%%%%%%%%%%%%%%%% %%%%%%%%%%%%%%%%%%%%%%%%%
%%%%%%%%%%%%%%%%%% motivate nHDMs %% %%%%%%%%%%%%%%%
%%%%%%%%%%%%%%%%%% %%%%%%%%%%%%%%%%%%%%%%%%%

The Standard Model (SM) of particle physics remains consistent in 
accommodating the experimental tests designed to measure its 
properties.  
The discovery of a scalar particle at the LHC has further vindicated the SM.  
This discovery has also intensified the interest in scalar extensions 
of the SM, which feature more than one fundamental scalars.  
In fact, the phenomenological evidences of dark matter 
and neutrino masses quite regularly motivate us to pursue physics beyond the 
SM (BSM). More often than not, these BSM scenarios come with an extension 
of the scalar sector of the SM.  
Although 
there are many different ways to extend the SM scalar sector,
extensions with additional $SU(2)_L$ doublets are particularly attractive because 
they preserve the tree-level value of the $\rho$-parameter~\cite{Workman:2022ynf}.
\iffalse,
%
\begin{equation}
\label{e:rho}
\rho = \frac{M_W^2}{M_Z^2 \cos^2\theta_W} = 1, 
\end{equation}
where $\theta_W$ is the weak mixing angle, and $M_{W}$ ($M_Z$) is the $W$ ($Z$) 
boson mass.  
\fi
%%%%%%%%%%%%%%%%%% %%%%%%%%%%%%%%%%%%%%%%%%%
%%%%%%%%%%%%%% motivate NFC and Democratic  %%%%%%%%%%%%%%%
%%%%%%%%%%%%%%%%%% %%%%%%%%%%%%%%%%%%%%%%%%%

The SM being reliant on the minimal scalar sector containing only one $SU(2)_L$ doublet, is free from flavour-changing neutral currents (FCNCs) at the tree-level.  
This feature is not guaranteed to be preserved when one extends the SM scalar sector.  
In fact, BSM with multiple scalar doublets, in general, lead to the presence of tree-level FCNCs mediated by neutral scalars.  
However, experimental data suggest that the FCNC processes are strongly suppressed~\cite{Workman:2022ynf}, which makes the absence of tree-level FCNCs a desirable property of any BSM scenario.  
A common way to achieve this is to impose a symmetry which ensures that fermions of a particular charge couple to only one scalar doublet.  
Consequently, the fermionic mass matrices and the corresponding Yukawa matrices are simultaneously diagonalizable, thereby preventing the appearance of scalar-mediated FCNCs at the tree-level, just as in the SM.  
Such a possibility, in the context of multi Higgs-doublet models, is known as natural flavour conservation (NFC)~\cite{Glashow:1976nt}.  
The two Higgs-doublet model (2HDM) entertains four types of flavour universal NFC models (type-I, type-II, type-X, and type-Y), which have been extensively studied in the literature~\cite{Branco:2011iw}.  
Beyond these four possibilities, there is one more attractive option where a particular scalar doublet is reserved exclusively for each type of massive fermion, and the up-type quarks, the down-type quarks, and the charged-leptons couple to separate dedicated scalar doublets.  
Quite clearly, such an arrangement of Yukawa couplings is impossible within the 2HDM framework and one needs at least three scalar doublets to achieve it.  
It should be mentioned that this particular composition of Yukawa couplings is commonly dubbed as `democratic'\cite{Cree:2011uy} or `type-Z'\cite{Akeroyd:2016ssd} Yukawa structure.  
In this paper, we choose to refer to this possibility as `democratic Yukawa' and subsequently, the three Higgs-doublet models (3HDMs) that feature a democratic Yukawa structure will be collectively called `democratic 3HDMs'.  
These democratic 3HDMs have received a lot of attention in the recent past.  
Theoretical constraints from unitarity and boundedness from below (BFB) have been studied in refs.~\cite{Bento:2022vsb, Boto:2022uwv}, the alignment limit in democratic 3HDMs is studied in refs.~\cite{Das:2019yad, Pilaftsis:2016erj}, and more recently, the phenomenological studies involving the flavour and Higgs data have been performed in refs.~\cite{Chakraborti:2021bpy, Boto:2021qgu}.

In this paper, we turn our attention to a couple of unexplored aspects of democratic 3HDMs, namely, the custodial limit and the possibility of `wrong-sign' Yukawa couplings.  
Keeping in mind the surging popularity of democratic 3HDMs, this study is quite timely and relevant.  
To highlight the importance of the custodial limit, we recall that in the SM, the custodial symmetry (CS) ensures $\rho=1$ at the tree-level.  
%In a pure $SU(2)$ theory, the custodial symmetry can be exact, but in the SM the $\rho$-parameter has corrections coming from the $U(1)_Y$ gauge couplings and Yukawa couplings.
The custodial symmetry is only an approximate symmetry of the SM since it is broken by the $U(1)_Y$ gauge coupling, as well as the Yukawa couplings~\cite{Willenbrock:2004hu}.  
Because of this, at the loop level, the $\rho$-parameter deviates slightly from unity and the deviation is quite accurately predicted by the SM.  
As it happens, the experimental measurement is compatible with this SM prediction, leaving very little room for new physics (NP) to give an extra contribution.  
Such NP contributions are sometimes conveniently expressed in terms of the $T$-parameter, which has the following experimental limit~\cite{Workman:2022ynf}
\begin{equation}
\Delta T = 0.03 \pm 0.12 \nonumber \,\,.
\end{equation}
One noteworthy aspect is that the SM scalar sector respects CS perfectly.  
However, this is no longer guaranteed once the scalar sector is extended.  
Therefore, it is expected that the additional scalars will give rise to extra contributions to the $T$-parameter.  
The limit on the $T$-parameter will place constraints on the NP contributions, sometimes requiring a fine-tuned scalar spectrum to keep the value under control.  
Thus, models with $n$ Higgs-doublets (nHDMs), although respecting $\rho=1$ at the tree-level, can potentially drive the $T$-parameter away from the experimental bounds, if the scalar masses are arbitrarily chosen~\cite{Pomarol:1993mu,Haber:2010bw, Grimus:2007if, Grimus:2008nb}.  
Therefore, it can be very attractive if we can systematically construct democratic 3HDMs which respect the CS in the scalar sector by design.  
Admittedly, such studies have been performed earlier in the context of nHDMs~\cite{Olaussen:2010aq, Nishi:2011gc, Solberg:2018aav,Grzadkowski:2010dj,Darvishi:2019dbh}, resulting in relations between the quartic parameters of the scalar potential.  
But, unlike the earlier studies, which directly implement the custodial symmetry in the scalar potential, our analysis conveniently starts with the scalar kinetic terms, following ref.~\cite{Kundu:2021pcg}.  
This alternative approach enables us to intuitively identify the different custodial multiplets and at the end, the conditions for CS in nHDMs are concisely expressed in a single equation, in terms of the physical masses and mixings of the scalar sector.  
Note that such a condition does not depend on the explicit structure of the scalar potential.  
Being related to the mass matrices of the scalar sector, the condition for respecting CS in nHDMs becomes quite easily implemented in practical analysis.  
As a simple cross-check, we will also show how the conditions in terms of the quartic parameters of the scalar sector in earlier references follow from this single condition in a straightforward manner.

The scalar extensions of the SM also face severe constraints from the measurements of the Higgs signal strengths~\cite{ATLAS:2021vrm}.  
For nHDMs, these constraints can be greatly alleviated by staying in the proximity of the `alignment limit'~\cite{Das:2019yad, Das:2015mwa, Gunion:2002zf, Bhattacharyya:2013rya, Pilaftsis:2016erj, BhupalDev:2014bir, Carena:2013ooa, Darvishi:2021txa}, where the lightest CP-even scalar has the same couplings as the SM Higgs boson at the tree-level.  
However, an intriguing possibility may arise if we keep in mind that the current Higgs data is not very sensitive to the sign of the down quark and charged lepton Yukawa.
%\footnote{The possibility of a wrong-sign $hZZ$ coupling has been considered in ref.~\cite{deLima:2021llm}.}
Such an exotic possibility can be accommodated in a 2HDM framework with \textit{e.g.} a type-II Yukawa structure and is quite well studied in the literature~\cite{Fontes:2014tga, Biswas:2015zgk, Ferreira:2014dya,Ferreira:2014qda}.  
In this paper, we want to point out that democratic 3HDMs can also accommodate this possibility, with much more freedom, due to the increased number of parameters.  
These possibilities should encourage our experimental colleagues to report the results
of the analysis of the Higgs data in the kappa framework~\cite{LHCHiggsCrossSectionWorkingGroup:2012nn, LHCHiggsCrossSectionWorkingGroup:2013rie} without any implicit
assumption on the sign of the kappas.

This article will be organized as follows.  
In \Sect{s:CS_defn} we lay down our methodology to study the CS starting from the
scalar kinetic terms. We then apply this in the case of the SM and recover the essential
features of CS in the SM. Later in this section, we extend our analysis to the nHDM case
and retrieve the 2HDM result as a special example. In \Sect{s:tparameter} we explicitly
demonstrate how the custodial limit neutralizes the constraint arising from the
electroweak $T$-parameter.
We define democratic 3HDMs in \Sect{s:democratic}, and present the custodial limit for the two usual incarnations in \Sect{s:CS_lim_Dem3HDMs}.  
Afterwards, in \Sect{s:WS} we investigate the possibility of wrong-sign Yukawa couplings in democratic 3HDMs.  
Finally, we summarize our findings in \Sect{s:summary}.

%%%%%%%%%%%%%%%%%%%%%%%%%%%%%%%%%%%%%%%%%%%%%%%%%%%%%%%%%%%%%
%%%%%%%%%%%%%%%%Custodial Symmetry%%%%%%%%%%%%%%%%%%%%%%%%%%%
%%%%%%%%%%%%%%%%%%%%%%%%%%%%%%%%%%%%%%%%%%%%%%%%%%%%%%%%%%%%%
\section{Custodial Symmetry in multi Higgs-doublet models}
\label{s:CS_defn}
The CS is an accidental global $SU(2)$ symmetry (hereafter denoted as $SU(2)_C$) which prevails even after the spontaneous breaking of the electroweak symmetry in the SM.  
In the case of the SM gauge group, $SU(2)_L \times U(1)_Y$, the CS is responsible for the value of the $\rho$-parameter to be equal to unity at the tree-level.  
In this paper, we follow the formulation of CS as in ref.~\cite{Kundu:2021pcg}, and confine ourselves to the $SU(2)_L$ part of the electroweak gauge symmetry, that is, we work in the limit where the $U(1)_Y$ gauge coupling goes to zero ($g'=0$).  
In this section, we will build our intuition first, by considering the simple example of the SM scalar sector.  
Then, we will extend our formalism to the case of a general nHDM and obtain conditions such that the scalar sector obeys the CS.

\subsection{Recap of the custodial symmetry in the SM}
\label{s:recap}
In the SM, there is a single complex scalar doublet, $\phi$, which drives the electroweak symmetry breaking (EWSB).  
The scalar Lagragian of the SM is given by
\begin{equation}
\mathscr{L}_{\rm scalar} = \left( D^\mu \phi\right)^\dagger\left( D_\mu \phi\right) - V\left(\phi\right), 
\end{equation}
where $V(\phi)$ is the scalar potential.   
In the limit $g'=0$, the gauge-covariant derivative for $\phi$ is given by
\begin{eqnarray}	
	\label{e:sm_cd}
	D_\mu \phi &=&\left(\partial_\mu+i g \frac{\tau_a}{2}W_\mu^a\right) \phi ,
\end{eqnarray}
where $g$ is the $SU(2)_L$ gauge coupling, $W^a_\mu$ are the $SU(2)_L$ gauge bosons, and $\tau_a$ are the  Pauli matrices.  
After the EWSB, the scalar doublet $\phi$ can be explicitly expressed in terms of the component fields, as follows
\begin{eqnarray}
	\label{e:sm_hd}
	\phi=\frac{1}{\sqrt{2}}\begin{pmatrix}
		\sqrt{2} \, \omega^+ \\
		v+h+i \zeta
	\end{pmatrix} \, , 
\end{eqnarray}
where $v$ is the vacuum expectation value (VEV).  
Subsequently, the scalar kinetic terms can be conveniently decomposed as\cite{Kundu:2021pcg}
\begin{eqnarray}
	\label{e:lgn}
	\mathscr{L}_{\rm{kin}}= (D^\mu \phi)^\dagger (D_\mu \phi)	=\mathscr{L}_{\rm{mass}} +\mathscr{L}_{\rm{quad}}+ \mathscr{L}_{\rm{mixed}}+\mathscr{L}_{\rm{deriv}}+\mathscr{L}_{\rm{cubic}}+\mathscr{L}_{\rm{quartic}} \, .
\end{eqnarray}
Collectively denoting the gauge bosons as $G_\mu^{a, b, ...}$ and the component scalar fields as $s_{i,j,...}$, the meaning of the individual terms introduced in the above equation are given below
\begin{eqnarray}
	\mathscr{L}_{\rm{mass}}:&&\text{these are the mass terms for the guage bosons of the form $v^2 {G_\mu^a}^\dagger G^{a\mu}$}\, ,\nonumber\\
	\mathscr{L}_{\rm{quad}}:&&\text{these are the kinetic terms of the component scalar fields, $(\partial^\mu s_i)^\dagger (\partial_\mu s_i)$}\, ,\nonumber\\
	\mathscr{L}_{\rm{mixed}}:&&\text{terms of the form} \, (\partial^\mu s_i)^\dagger  (iv G_\mu)  \text{ + h.c.}\, ,\nonumber\\
	\mathscr{L}_{\rm{deriv}}:&&\text{terms of the form} \, (\partial^\mu s_i)^\dagger (iG_\mu s_j) \text{ + h.c.} \, ,\nonumber\\
	\mathscr{L}_{\rm{cubic}}:&&\text{terms of the form} \, (G^{a,\mu} s_i)^\dagger (vG^b_\mu ) \text{ + h.c.}\, ,\nonumber\\
	\mathscr{L}_{\rm{quartic}}:&&\text{terms of the form} \, (G^{a,\mu} s_i)^\dagger(G^b_\mu s_j)\, .\nonumber
\end{eqnarray}
To identify the custodial multiplets, we begin with $\mathscr{L}_{\rm mass}$ which, in the SM, is given by
\begin{eqnarray}
	\label{e:sm_lmass1}
	\mathscr{L}_{\rm{mass}} &=& \frac{g^2 v^2}{8}\left(W^{\mu+}W_\mu^-  + W^{\mu-}W_\mu^+ + W^{3  \mu}W_\mu^3  \right) \, .
\end{eqnarray}
where 
\begin{equation}
W_\mu^{ \pm} = \dfrac{W_\mu^{1} \mp i W_\mu^{2}}{\sqrt{2}}. 
\end{equation}
We can see from the above equation that the $SU(2)_L$ gauge bosons have the same mass.  
This motivates us to identify a custodial multiplet of the gauge bosons as\footnote{the minus sign in the first entry of $\mathbf{W}$ comes from the details of $SU(2)$ group theory, which are explained in Appendix~\ref{s:SU(2)GT}.} 
\begin{eqnarray}
	\label{e:sm_W}
	\textbf{W}=\begin{pmatrix}
		-W^+\\W_3\\W^-
	\end{pmatrix} \,.
\end{eqnarray}
Note that the Lorentz indices have been suppressed here for simplicity as it has no bearing on the $SU(2)_C$ transformations.  
In terms of the $SU(2)_C$ triplet of Eq.~\eqref{e:sm_W}, $\mathscr{L}_{\rm mass}$ can be rewritten as 
\begin{eqnarray}
	\label{e:sm_lmass}
	\mathscr{L}_{\rm{mass}}=\frac{g^2 v^2}{8}\left(\textbf{W}\cdot\textbf{W}\right),
\end{eqnarray}
which is manifestly invariant under $SU(2)_C$.  
To identify the $SU(2)_C$ multiplets of the scalar fields, let us turn our attention to $\mathscr{L}_{\rm cubic}$ and $\mathscr{L}_{\rm mixed}$.  
First, in terms of the triplet $\mathbf{W}$, $\mathscr{L}_{\rm cubic}$ can be expressed as
\begin{eqnarray}
	\label{e:sm_lcubic}
	\mathscr{L}_{\rm{cubic}} = \frac{g^2v}{4}h\left(\textbf{W}\cdot\textbf{W}\right) \, .
\end{eqnarray}
Thus, $\mathscr{L}_{\rm cubic}$ will also be $SU(2)_C$ invariant if we identify the physical scalar, $h$, as a singlet of $SU(2)_C$.  
Next, we look into $\mathscr{L}_{\rm mixed}$, which is given by
\begin{eqnarray}
	\label{e:sm_lmix1}
	\mathscr{L}_{\rm{mixed}}=  \frac{gv}{2}   \bigg[i\left(\partial^\mu w^-\right) W_\mu^+  -  i\left(\partial^\mu w^+\right)W_\mu^-  -  \left(\partial^\mu \zeta\right) W_\mu^3 \bigg]\, .
\end{eqnarray}
Given the identification of $\mathbf{W}$ in Eq.~\eqref{e:sm_W}, the above equation encourages us to define an $SU(2)_C$ triplet of scalar fields as follows:
\begin{eqnarray}
	\label{e:sm_T}
	\textbf{T}=\begin{pmatrix}
		i\omega^+\\-\zeta\\i\omega^-
	\end{pmatrix}\,.
\end{eqnarray}
In terms of $\mathbf{W}$ and $\mathbf{T}$, Eq.~\eqref{e:sm_lmix1} can be written as
\begin{eqnarray}
	\label{e:sm_lmix}
	\mathscr{L}_{\rm{mixed}}=\frac{g v}{2}(\textbf{W}\cdot \partial\textbf{T})\, ,
\end{eqnarray}
which explicitly demonstrates the $SU(2)_C$ invariance of $\mathscr{L}_{\rm mixed}$.  
The other terms, $\mathscr{L}_{\rm quad}$, $\mathscr{L}_{\rm deriv}$, and $\mathscr{L}_{\rm quartic}$, when expressed in terms of $\mathbf{W}$ and $\mathbf{T}$, can also be shown to be invariant under $SU(2)_C$.  
All these terms will be considered in detail in the next subsection, when we consider the nHDM generalisation of the above prescription.  

Now, let us take a look at the $SU(2)_C$ invariance of the scalar potential, which is given by
\begin{eqnarray}
	\label{e:sm_pot}
	V\left(\phi\right)=\mu^2\left(\phi^\dagger \phi\right)+\lambda\left(\phi^\dagger \phi\right)^2 \, .
\end{eqnarray}
After the EWSB, $\phi^\dagger \phi$ can be expressed as 
\begin{eqnarray}
	\label{e:sm_pot1}
	\phi^\dagger \phi&=&\frac{1}{2}(\mathbf{T}\cdot\mathbf{T})+\frac{v^2}{2}+\frac{h^2}{2}+vh\,.
\end{eqnarray}
We can see that, our previous multiplet identifications of $\textbf{T}$ and $h$ are compatible with the $SU(2)_C$ invariance of the scalar potential.  
In other words, no additional conditions need to be imposed on the SM scalar potential to make it $SU(2)_C$ invariant.  
It should be noted that, the $SU(2)_C$ invariance of the scalar potential mandates that the scalars which are in the same $SU(2)_C$ multiplet should have the same mass.  
This condition is trivially satisfied here in the SM as all the components of $\mathbf{T}$ are Goldstone bosons with zero masses.  
This will no longer be true in nHDMs, where we will need to impose additional restrictions on the parameters of the scalar potential to ensure custodial invariance.

%%%%%%%%%%%%%%%%%%%%%%%%%%%%%%%%%%%%%%%%%%%%%%%%%%%%
%%%%%%%%%%%%%%%%%%%%%%%% L_kinetic nHDM
%%%%%%%%%%%%%%%%%%%%%%%%%%%%%%%%%%%%%%%%%%%%%%%%%%%%%%%5
\subsection{Generalization to nHDM }
\label{s:nHDMs}
We will now look at the scalar kinetic Lagrangian for a model with $n$ complex scalar doublets $\phi_k$ ($k=1, \dots\,, n$) and identify the different $SU(2)_C$ multiplets.  
Thus we begin with
\begin{eqnarray}
	\label{e:lkin}
	\mathscr{L}_{\rm{kin}}= \sum_{k=1}^n(D^\mu\phi_k)^\dagger(D_\mu\phi_k)\,,
\end{eqnarray}
where, under the assumption of $g'=0$, the gauge covariant derivative of $\phi_k$ is given by
\begin{eqnarray}
	D_\mu \phi_k = \left(\partial_\mu  + ig\frac{\tau_a}{2}W_\mu^a\right) \phi_k\,.
\end{eqnarray}
After the EWSB, the $k$-th scalar doublet is decomposed as
\begin{eqnarray}
	\label{e:hd}
	\phi_k=\frac{1}{\sqrt{2}}\begin{pmatrix}
		\sqrt{2} w_k^+ \\
		v_k+h_k+i z_k
	\end{pmatrix} \, ,
\end{eqnarray}
where $v_k$ is the VEV of $\phi_k$, assumed to be real.  
Borrowing the terminology introduced in Eq.~\eqref{e:lgn}, we still have 
\begin{eqnarray}
	\label{e:lmass1}
	\mathscr{L}_{\rm{mass}}=\frac{g^2 v^2}{8}\left(\textbf{W}\cdot\textbf{W}\right),
\end{eqnarray}
where $v= \sqrt{v_1^2+v_2^2+...+v_n^2}$ is the total electroweak VEV, and we have used Eq.~\eqref{e:sm_W} for the definiton of $\mathbf{W}$.  
This implies that $\mathscr{L}_{\rm mass}$ will still respect $SU(2)_C$ once we identify the custodial triplet of the gauge bosons, as in the case of the SM.  
Similarly, for $\lag{cubic}$ we have
\begin{eqnarray}
	\label{e:nlcubic}
	\mathscr{L}_{\rm{cubic}}= \frac{g^2}{4} \left( \mathbf{W} \cdot \mathbf{W} \right) \sum_{k=1}^{n} v_k h_k \, .
\end{eqnarray}
Evidently, $\lag{cubic}$ will also be custodially invariant if we identify $h_k$ ($k=1, ..., n$) as singlets of $SU(2)_C$.  
Next, we turn our attention to $\lag{mixed}$, which has the following form
\begin{eqnarray}
	\label{e:Lmix1}
	\mathscr{L}_{\rm{mixed} }=  \frac{g}{2} \sum_{k=1}^{n}v_k   \left[i(\partial^\mu w_k^-) W_\mu^+  -  i(\partial^\mu w_k^+)W_\mu^-  -  (\partial^\mu z_k) W_\mu^3 \right] \, .
\end{eqnarray} 
Taking inspiration from Eq.~\eqref{e:sm_lmix1}, we now proceed to define a set of $SU(2)_C$ triplets involving the scalar component fields as 
\begin{eqnarray}
	\label{e:ntriplet}
	\textbf{T}_k \equiv \begin{pmatrix} 
		iw_k^+\\
		-z_k\\
		iw_k^-\end{pmatrix}\,, \quad k=1, ...,n \, .
\end{eqnarray}
Following this identification, we can express $\lag{mixed}$ as the sum of $SU(2)_C$ invariants, given by
\begin{eqnarray}
	\label{e:NLmix}
	\mathscr{L}_{\rm{mixed}} &=& \frac{g}{2} \sum_{k=1}^{n} v_k \left(\mathbf{W}\cdot\partial\mathbf{T}_k\right)\,.
\end{eqnarray}
For the sake of completeness, we also express $\lag{quad}$, $\lag{quartic}$, and $\lag{deriv}$, in terms of $\mathbf{W}$, $\mathbf{T}_k$, and $h_k$,  as follows
\begin{subequations}
\begin{eqnarray}
	\label{e:nlquad}
	\mathscr{L}_{\rm{quad}} &=& \frac{1}{2} \sum_{k=1}^{n}\left[(\partial \mathbf{T}_k\cdot \partial \mathbf{T}_k)+(\partial^\mu h_k)( \partial_\mu h_k)\right], \\
	\label{e:nlqu}
	\mathscr{L}_{\rm{quartic}} &=& \frac{g^2}{8} (\textbf{W}\cdot\textbf{W}) \sum_{k=1}^{n} (\textbf{T}_k\cdot\textbf{T}_k + h_k^2)\,, \\
	\label{e:nlder}
	\mathscr{L}_{\rm{deriv}} &=& \frac{g}{2} \sum_{k=1}^{n} \left\{h_k(\textbf{W}\cdot \partial \textbf{T}_k  ) + \partial h_k( \textbf{T}_k\cdot \textbf{W}) + (\textbf{T}_k\times \partial \textbf{T}_k )\cdot\textbf{W}\right\}\,,
\end{eqnarray}
\end{subequations}
where $(\mathbf{r}_1 \times \mathbf{r}_2 ) \cdot \mathbf{r}_3$ is the singlet combination of the $SU(2)$ product of three triplets, $\mathbf{r}_{1,2,3}$, for which the explicit expression is given in appendix~\ref{s:SU(2)GT}.  

Thus, we can see that all the terms in the scalar kinetic Lagrangian are custodially invariant.  
However, the triplets $\mathbf{T}_k$ are not expressed in terms of physical fields.  
Rotation of these fields from the Lagrangian basis to the physical basis will give rise to the Goldstone bosons, the physical charged scalars, and pseudoscalars\footnote{Following Ref.~\cite{Nishi:2011gc}, it is reasonable to have  such a classification of the scalar spectrum because CP conservation follows for nHDMs with custodial symmetry.} 
We would like to transfer the $SU(2)_C$ invariance into the physical basis as well.  
For this, we need to rotate each triplet as a whole object, that is,
\begin{eqnarray}
	\label{e:ntripletrot}
	\mathbf{P}_j= \sum_{k=1}^{n} \mathcal{O}_{jk}\mathbf{T}_k \quad \quad j=1,2,\dots n \,,
\end{eqnarray}
where $\mathbf{P}_j$ denotes the $j$-th triplet of $SU(2)_C$ in the physical basis, and $\mathcal{O}_{jk}$ are the elements of an orthogonal matrix.  
Note that, each triplet $\mathbf{T}_k$, contained a pseudoscalar field and a pair of charged fields.  
Consequently, Eq.~\eqref{e:ntripletrot} implies that the charged and pseudoscalar mass matrices should be rotated into the physical basis by means of the same rotation matrix, in order to preserve the $SU(2)_C$ invariance of $\lag{kin}$ in the physical basis as well.  
Now, for a charged scalar and a pseudoscalar in the physical basis to be placed in the same triplet $\mathbf{P}_j$, they should have a common mass so that the mass terms for the members of $\mathbf{P}_j$ can be concisely expressed in an $SU(2)_C$ invariant form as $M_j^2 \, (\mathbf{P}_j \cdot \mathbf{P}_j)$.  
Thus, we can conclude that, in the physical basis, the diagonal mass matrices in the charged and pseudoscalar sectors must be equal.  
Also, from Eq.~\eqref{e:ntripletrot}, we should recall that the rotations that bring the mass matrices of the charged and pseudoscalar sectors to their respective diagonal forms should also be the same.  
Putting this together, we can conclude that the mass matrix of the charged and pseudoscalar sectors should be equal in the Lagrangian basis as well, that is
\begin{eqnarray}
	\label{e:mceqmp}
	M_{C}^2=M_{P}^2 \,.
\end{eqnarray}
Since the information about the scalar masses and the mixings comes from the scalar potential, the parameters of the scalar potential should adjust themselves so that Eq.~\eqref{e:mceqmp} is satisfied for arbitrary values of the VEVs. 
The arbitrariness of the VEVs is important because the validity of the custodial symmetry should not depend on the exact values of the VEVs, just as in the case of the SM.
%\revision{This is a consequence of $SU(2)_C$ not being spontaneously broken by the EWSB, thus requiring the custodial invariance condition to hold for all values of the VEVs. Consequently, Eq.~\eqref{e:mceqmp} (rather than a equality) is a statement about the quartic and bilinear parameters of the scalar potential, such that it also comes under the umbrella of $SU(2)_C$ invariance.  }

\subsection{Examples with 2HDMs}
We will now explicitly demonstrate how Eq.~\eqref{e:mceqmp} manifests itself for the simple case of a 2HDM scalar potential.  
At first, let us consider the 2HDM scalar potential with a softly-broken $Z_2$ symmetry ($\phi_1\to\phi_1, \phi_2\to-\phi_2$), which is commonly used in NFC models~\cite{Branco:2011iw}:
\begin{eqnarray}
	\label{e:2scapot}
	V(\phi_1, \phi_2) &=& m_{11}^2 \phi_1^\dagger \phi_1 + m_{22}^2\phi_2^\dagger \phi_2 - m_{12}^2( \phi_1^\dagger \phi_2 + \phi_2^\dagger \phi_1) + \frac{\lambda_1}{2}(\phi_1^\dagger \phi_1)^2 + \frac{\lambda_2}{2}(\phi_2^\dagger \phi_2)^2 \nonumber \\
	&& + \lambda_3 (\phi_1^\dagger \phi_1)(\phi_2^\dagger \phi_2) + \lambda_4(\phi_1^\dagger \phi_2)(\phi_2^\dagger \phi_1) + \frac{\lambda_5}{2} \left\{  (\phi_1^\dagger \phi_2)^2  + (\phi_2^\dagger \phi_1)^2  \right\} .
\end{eqnarray}
The charged and pseudoscalar mass matrices which transpire from the above scalar potential are given by
\begin{subequations}
\begin{eqnarray}
	\label{e:2mcsq}
	M_C^2 &=& \begin{pmatrix}
	\frac{m_{12}^2 v_2}{v_1} -\frac{1}{2}\lambda_4 v_2^2  -\frac{1}{2}\lambda_5 v_2^2 & -m_{12}^2 + \frac{1}{2} \lambda_4 v_1 v_2 + \frac{1}{2} \lambda_5 v_1 v_2 \\
	-m_{12}^2 + \frac{1}{2}\lambda_4 v_1 v_2 + \frac{1}{2}\lambda_5 v_1 v_2 & \frac{m_{12}^2 v_1}{v_2} -\frac{1}{2} \lambda_4 v_1^2 -\frac{1}{2} \lambda_5 v_1^2
	\end{pmatrix}\,, \\
	\label{e:2mpsq}
	M_P^2 &=& \begin{pmatrix}
	\frac{m_{12}^2 v_2}{v_1} - \lambda_5 v_2^2 & - m_{12}^2+ \lambda_5 v_1v_2  \\
	- m_{12}^2+ \lambda_5 v_1v_2 & \frac{m_{12}^2 v_1}{v_2} -\lambda_5 v_1^2
	\end{pmatrix}\,.
\end{eqnarray}
\end{subequations}
Thus, imposition of Eq.~\eqref{e:mceqmp} for arbitrary values of the VEVs will lead to the following relation
\begin{eqnarray}
	\label{e:z2_conditions}
	\lambda_4=\lambda_5\, ,
\end{eqnarray}
which agrees with earlier results~\cite{Pomarol:1993mu,Haber:2010bw,Darvishi:2019dbh}.
In passing, we wish to point out that even if we consider the general 2HDM potential~\cite{Branco:2011iw} %\textcolor{blue}{Rewrite}
\begin{eqnarray}
	\label{e:2hdmpot1}
	V(\phi_1, \phi_2) &=& m_{11}^2 \phi_1^\dagger \phi_1 + m_{22}^2\phi_2^\dagger \phi_2 - (m_{12}^2 \phi_1^\dagger \phi_2 + \text{h.c.}) + \frac{\lambda_1}{2}(\phi_1^\dagger \phi_1)^2 + \frac{\lambda_2}{2}(\phi_2^\dagger \phi_2)^2+ \lambda_3 (\phi_1^\dagger \phi_1)(\phi_2^\dagger \phi_2)  \nonumber \\
	&& + \lambda_4(\phi_1^\dagger \phi_2)(\phi_2^\dagger \phi_1) + \left\{ \frac{\lambda_5}{2} (\phi_1^\dagger \phi_2)^2 +\lambda_6(\phi_1^\dagger \phi_1)(\phi_1^\dagger \phi_2)+\lambda_7(\phi_2^\dagger \phi_2)(\phi_1^\dagger \phi_2)+\text{h.c.} \right\}\, , 
\end{eqnarray}
the condition for custodial invariance is still given by  Eq.~\eqref{e:z2_conditions}.  
The reason for this will be discussed in more detail in Appendix~\ref{s:compare}.

%%%%%%%%%%%%%%%%%%%%%%%%%%%%%%%%%%%%%%%%%%%%%%%%%%%%%%%%%
%%%%%%%%%% T parameter   %%%%%%%%%%%%%%%%%%%%%
%%%%%%%%%%%%%%%%%%%%%%%%%%%%%%%%%%%%%%%%%%%%%%%%%%%%%%%%
\section{Validation of the custodial limit by explicit calculation}
\label{s:tparameter}
In $SU(2)_C$ invariant models, we expect that no additional contribution to the $T$-parameter comes from the scalar sector.  
It would be rather reassuring to explicitly verify that this is indeed the case for nHDMs in the limit of Eq.~\eqref{e:mceqmp}. For this purpose,  we use the one-loop formula for the NP contribution to the $T$-parameter for nHDMs given in refs.~\cite{Grimus:2007if, Grimus:2008nb}:  

\begin{eqnarray}
\label{e:grimus_v2}
\alpha T = \frac{g^2}{64\pi^2M_W^2} &\Bigg\{& \sum_{a=2}^{n} \sum_{b=2}^{2n} \left\lvert \big( U^\dagger V \big)_{ab} \right\rvert^2F\left(m_a^2,\mu_b^2\right) - \sum_{b=2}^{2n-1}\sum_{b'=b+1}^{2n}\left\lvert\big(V^\dagger V\big)_{bb'}\right\rvert^2 F\left(\mu_b^2,\mu_{b'}^2\right) \nonumber \\&&  +3\sum_{b=2}^{n}\left\lvert\big(V^\dagger V\big)_{1b}\right\rvert^2\bigg[F\left(M_Z^2,\mu_b^2\right)-F\left(M_W^2,\mu_b^2\right)\bigg]\Bigg\}\,,
\end{eqnarray}
where 
\begin{eqnarray}
\label{e:tparam_Fxy_v2}
F(x,y) \equiv \left\{ \begin{matrix} \dfrac{x+y}{2}-\dfrac{xy}{x-y}\ln\dfrac{x}{y}, & x\neq y \\ 0, &  x=y \end{matrix} \right. \, ,
\end{eqnarray}
and $\alpha$ is the fine-structure constant.  
The masses of the charged-scalars are denoted by $m_a$, and $\mu_a$ are the masses of the physical neutral scalars, defined in such a way that $a \leq n$ refers to the pseudoscalars, and $a >  n$ are the CP-even fields.  
Lastly, $U^\dagger$ and $V^\dagger$ are $n\times n$ and $2 n \times n$ matrices that rotate the charged and neutral components ($w_k^\pm$ and $\varphi_k^0 \equiv h_k + i z_k$) into the physical basis ($S^\pm$ and $S^0$), respectively, in such a way that the Goldstone bosons are located in the first row, 
\begin{eqnarray}
\label{eq:basisrot}
w_k^\pm = \sum_{a=1}^{n} U_{ka} S_a^\pm, \qquad \varphi_k^0 = \sum_{b=1}^{2n} V_{kb} S_b^0\, .
\end{eqnarray}
We give the explicit structure of $S^\pm$ and $S^0$ as follows
\begin{eqnarray}
S^\pm = \left( \omega^\pm ,  H_1^\pm  ,  \dots  ,  H_{n-1}^\pm \right)^T \, , \qquad S^0 = \big( \zeta ,  A_1  ,  \dots  ,  A_{n-1},  h,  H_1, \dots ,  H_{n-1} \big)^T , 
\end{eqnarray}
where $\omega^\pm$ and $\zeta$ are the charged and neutral unphysical Goldstone bosons, respectively,  
$H_k^\pm$ is the $k$-th charged scalar, and $A_k$ the $k$-th pseudoscalar.  
For the CP-even scalars, $h$ is the lightest scalar usually identified as the SM-like Higgs, and $H_k$ denotes the $k$-th physical CP-even scalar.  
Following the definition of Eq.~\eqref{e:ntripletrot}, and comparing with Eq.~\eqref{eq:basisrot}, we can relate the $U$ and $V$ matrices with the scalar rotation matrices as follows
\begin{eqnarray}
\label{e:tparam_rot12}
U=\mathcal{O}_C^T \,, \quad 
V=\begin{pmatrix}
i\mathcal{O}_P^T&\mathcal{O}_S^T
\end{pmatrix}, 
\end{eqnarray} 
where the subscripts $C, P, S$ refer to the charged, pseudoscalar, and scalar sectors respectively.  
The relevant combinations can be expressed as
\begin{eqnarray}
\label{e:rotsquareds}
U^\dagger V = \begin{pmatrix} i \, \mathcal{O}_C \, \mathcal{O}_P^T \,\,  & \mathcal{O}_C \, \mathcal{O}_S^T \end{pmatrix}, \qquad 
V^\dagger V = \begin{pmatrix} \mathbb{1}_{n\times n} \,\, & -i \, \mathcal{O}_P \, \mathcal{O}_S^T \\ i \, \mathcal{O}_S \, \mathcal{O}_P^T \,\,  & \mathbb{1}_{n \times n}\end{pmatrix}.
\end{eqnarray}
We must note that the last term of Eq.~\eqref{e:grimus_v2} vanishes in the limit $g'\to0$, that is, $M_Z=M_W$.  
Therefore, we will focus on the first two terms in Eq.~\eqref{e:grimus_v2}, and convince ourselves that they also vanish in the custodial limit of Eq.~\eqref{e:mceqmp}.  
Taking advantage of Eq.~\eqref{e:rotsquareds}, we can rewrite the first two terms of Eq.~\eqref{e:grimus_v2} as 
\begin{eqnarray}
\label{e:grimus_v3}
\alpha T =\frac{g^2} {64\pi^2M_W^2} &\bigg\{ &
\sum_{a=2}^{n}\sum_{b=2}^{n} \left\lvert \big( i \mathcal{O}_C \mathcal{O}_P^T \big)_{ab} \right\rvert^2 F \left( m_a^2, \mu_b^2 \right) + \sum_{a=2}^{n}\sum_{b=1}^{n} \left\lvert \big( \mathcal{O}_C \mathcal{O}_S^T \big)_{ab}\right\rvert^2 F\left( m_a^2, \mu_{n+b}^2\right) \nonumber \\
&& - \sum_{a=2}^{n}\sum_{b=1}^{n} \left\lvert \big(-i \mathcal{O}_P \mathcal{O}_S^T \big) _{ab} \right\rvert^2 F\left( \mu_a^2, \mu^2_{n+b} \right) \bigg\}.
\end{eqnarray}
In the custodial limit, we must have $M_P^2=M_C^2$, and thus $\mathcal{O}_P = \mathcal{O}_C$, as well as $m_a^2 = \mu_a^2$ (with $a <n$).  
In this way, the second and third terms of Eq.~\eqref{e:grimus_v3} cancel out, and $\mathcal{O}_C \mathcal{O}_P^T = \mathcal{O}_P \mathcal{O}_P^T = \mathbb{1}_{n \times n}$ leads to a zero contribution from the first term, because of Eq.~\eqref{e:tparam_Fxy_v2}.
%

%%%%%%%%%%%%%%%%%%%%%%%%%%%%%%%%%%%%%%%%%%%
%%%%%%%%%%% 3HDMS
%%%%%%%%%%%%%%%%%%%%%%%%%%%%%%%%%%%%%%%%%%%
\section{Democratic 3HDMs}
\label{s:democratic}
%
%Natural Flavor Conservation in the scalar sector is inbuilt into the SM. This is because in the mass basis of the fermions, the Yukwawa matrices are diagonal thus eliminating flavor violating interactions. However, adding more Higgs doublets can destroy this feature if we allow more than one Higgs doublet to couple to the same fermions. Democratic nHDMs are models constructed to preserve NFC. This is done by imposing additional symmetries on the Lagrangian such that each fermion only couples to one Higgs doublet. 3HDMs are the minimal choice to have a democratic structure wherein up-type quarks, down-type quarks and leptons each couple to their own designated Higgs doublet. The Yukawa Lagrangian for a democratic 3HDM has the following form

The Yukawa Lagrangian for a democratic 3HDM, as discussed in the introduction, has the following form
\begin{eqnarray}
	\label{e:d3hdmlyuk}
	\mathscr{L}_Y= - Y_d \bar{Q}_L\phi_2 n_R - Y_u \bar{Q}_L\tilde{\phi}_3 p_R - Y_\ell \bar{L }_L \phi_1 \ell_R \, ,
\end{eqnarray}
where $Y_{d.u.\ell}$ are the Yukawa couplings in the down-quark, up-quark, and charged-lepton sectors.  
The up-type, down-type, and charged-lepton right-handed fields are denoted as $p_R$, $n_R$, and $\ell_R$, respectively.  
The left-handed $SU(2)_L$ doublets for the quarks and leptons are $Q_L = \left( p_L, \, n_L\right)^T$ and $L_L = \left( \nu_L, \, e_L\right)^T$.  
Finally, $\tilde{\phi}_3 = i \tau_2 \phi_3^*$ is the $SU(2)_L$ doublet responsible for the up-quark masses.
There are two common ways to arrive at the above Lagrangian.  
The first is to impose a $Z_3$ symmetry as follows~\cite{Das:2019yad}
\begin{eqnarray}
\label{e:z3assignments}
\phi_1 \to \omega \, \phi_1 \,, \qquad  \phi_2 \to \omega^2 \phi_2 \,,  \qquad	\ell_R \to \omega^2 \ell_R \,,  \qquad  n_R \to \omega \,n_R \,.  
\end{eqnarray}
The second possibility relies on a $Z_2 \times Z_2'$ symmetry under which the fields transform as~\cite{Akeroyd:2016ssd}
\begin{subequations}
\label{e:z2xz2-defn}
\begin{eqnarray}
	\label{e:z2z2_scalars}
	Z_2:&&\phi_1 \to -\phi_1 \, , \quad  \ell_R \to -\ell_R \\
	Z'_2:&&\phi_2 \to -\phi_2 \, , \quad  n_R \to -n_R 
\end{eqnarray}
\end{subequations}
Both in Eqs.~\eqref{e:z3assignments} and \eqref{e:z2xz2-defn}, only the nontrivial transformations are explicitly displayed.
In the following, we will discuss the implications of these symmetries on the scalar potential in the context of the custodial limit.  

%%%%%%%%%%%%%%%%%%%%%%%%%%%%%%%%%%%%%%%%%%%%%%%%%%%%%%%%%%%%%%%%%%%%%%%
\subsection{Custodial Limit of Democratic 3HDMs}
\label{s:CS_lim_Dem3HDMs}
In this subsection, we will write down the explicit forms of the scalar potential which follow from the symmetry of Eqs.~\eqref{e:z3assignments} and~\eqref{e:z2xz2-defn}.  
Then, we will proceed to calculate the detailed structure of the charged and pseudoscalar mass matrices.   
Finally, we will impose Eq.~\eqref{e:mceqmp} to extract the implications in terms of the parameters of the scalar potential.

\subsubsection{The case with a $Z_3$ symmetry}
\label{ss:Z3}

The scalar potential for this case will be given by~\cite{Bento:2017eti}
{\small
\begin{eqnarray}
	\label{e:z3_pot_v2}
	V_{\rm{Z_3}} &=& 
	 m_{11}^2 \phi_1^\dagger \phi_1 + m_{22}^2 \phi_2^\dagger \phi_2 + m_{33}^2 \phi_3^\dagger \phi_3 
	- m_{12}^2 (\phi_1^\dagger \phi_2 + \phi_2^\dagger \phi_1) - m_{13}^2 (\phi_1^\dagger \phi_3 + \phi_3^\dagger \phi_1 ) -m_{23}^2 (\phi_2^\dagger \phi_3 + \phi_3^\dagger \phi_2)  \nonumber\\
	&&+ \lambda_1 (\phi_1^\dagger \phi_1 )^2 
	+\lambda_2 (\phi_2^\dagger \phi_2 )^2
	+ \lambda_3 (\phi_3^\dagger \phi_3 )^2+ 
	 \lambda_4 (\phi_1^\dagger \phi_1 ) (\phi_2^\dagger \phi_2 ) 
	 +\lambda_5\ (\phi_1^\dagger \phi_1 ) (\phi_3^\dagger \phi_3 ) 
	 + \lambda_6 (\phi_2^\dagger \phi_2 ) (\phi_3^\dagger \phi_3 )  \nonumber \\ 
	&& 	+ \lambda_7 (\phi_1^\dagger \phi_2 ) (\phi_2^\dagger \phi_1 ) 
	+ \lambda_8 (\phi_1^\dagger \phi_3 ) (\phi_3^\dagger \phi_1 ) 
	 +\lambda_9 (\phi_2^\dagger \phi_3 ) (\phi_3^\dagger \phi_2 ) 
	 + \lambda_{10}\left\{ (\phi_1^\dagger \phi_2 ) (\phi_1^\dagger \phi_3 )+ (\phi_2^\dagger \phi_1 ) (\phi_3^\dagger \phi_1 ) \right\} 	  \nonumber\\
	   && + \lambda_{11}\left\{ (\phi_2^\dagger \phi_1 ) (\phi_2^\dagger \phi_3 )+ (\phi_1^\dagger \phi_2 ) (\phi_3^\dagger \phi_2 ) \right\} 
	   + \lambda_{12}\left\{ (\phi_3^\dagger \phi_1 ) (\phi_3^\dagger \phi_2 )+ (\phi_1^\dagger \phi_3 ) (\phi_2^\dagger \phi_3 ) \right\}\,,
\end{eqnarray}
}
where soft-breaking terms have also been allowed.  
The explicit expressions for the elements of the $3\times3$ symmetric mass matrix in the charged scalar sector are given below\footnote{We have used the minimization conditions to trade $m_{11}^2$, $m_{22}^2$, and $m_{33}^2$ in favor of the VEVs.}
\begin{subequations}
	\begin{eqnarray}
	\label{e:Mcsqz3}
	(M_C^2)_{11} &=&   \frac{ m_{12}^2 v_2}{v_1} + \frac{ m_{13}^2 v_3}{v_1} - \lambda_{10} v_2 v_3-\frac{\lambda_{11} v_2^2 v_3}{2 v_1} -\frac{\lambda_{12} v_2 v_3^2}{2 v_1}  -\frac{\lambda_7 v_2^2}{2} -\frac{\lambda_8 v_3^2}{2} \,,\\
	(M_C^2)_{22} &=&   \frac{m_{12}^2 v_1}{v_2} + \frac{m_{23}^2 v_3}{v_2} - \frac{\lambda_{10} v_1^2 v_3}{2v_2} - \lambda_{11} v_1 v_3 - \frac{\lambda_{12} v_1 v_3^2}{2 v_2} - \frac{\lambda_7 v_1^2}{2}  - \frac{\lambda_9 v_3^2}{2} \,,\\
	(M_C^2)_{33} &=& \frac{ m_{13}^2 v_1}{v_3} + \frac{m_{23}^2 v_2}{v_3} - \frac{\lambda_{10} v_1^2 v_2}{2 v_3} - \frac{\lambda_{11} v_1 v_2^2}{2 v_3} - \lambda_{12} v_1 v_2   - \frac{\lambda_8 v_1^2}{2} - \frac{\lambda_9 v_2^2}{2} \,,\\
	(M_C^2)_{12}    &=&  	(M_C^2)_{21}   = - m_{12}^2 + \frac{1}{2}\lambda_{10}v_1 v_3 + \frac{1}{2}\lambda_{11}v_2 v_3  + \frac{1}{2}\lambda_7 v_1 v_2 \,,\\
	(M_C^2)_{13}     &=&  	(M_C^2)_{31}    =  - m_{13}^2 + \frac{1}{2} \lambda_{10} v_1 v_2+ \frac{1}{2} \lambda_{12} v_2 v_3+ \frac{1}{2} \lambda_8 v_1 v_3 \,,\\
	(M_C^2)_{23}    &=&  	(M_C^2)_{32}  = - m_{23}^2 + \frac{1}{2} \lambda_{11} v_1 v_2 + \frac{1}{2} \lambda_{12} v_1 v_3 + \frac{1}{2} \lambda_9 v_2 v_3 \,.
	\end{eqnarray}
\end{subequations}
Similarly, for the pseudoscalar mass matrix we have
\begin{subequations}
	\begin{eqnarray}
	\label{e:Mpsqz3}
	(M_P^2)_{11}    &=&  \frac{ m_{12}^2 v_2}{v_1} +\frac{ m_{13}^2 v_3}{v_1} -2 \lambda_{10} v_2 v_3 -\frac{\lambda_{11} v_2^2 v_3}{2 v_1} -\frac{\lambda_{12} v_2 v_3^2}{2 v_1} \,, \\ 
	(M_P^2)_{22}  &=& \frac{m_{12}^2 v_1}{v_2} +\frac{m_{23}^2 v_3}{v_2} -\frac{\lambda_{10} v_1^2 v_3}{2 v_2} -2 \lambda_{11} v_1 v_3 -\frac{\lambda_{12} v_1 v_3^2}{2 v_2} \,,\\
	(M_P^2)_{33}   &=&  \frac{m_{13}^2 v_1}{v_3} +\frac{ m_{23}^2 v_2}{v_3} -\frac{\lambda_{10} v_1^2 v_2}{2 v_3} -\frac{\lambda_{11} v_1 v_2^2}{2 v_3} -2 \lambda_{12} v_1 v_2  \,,\\
	(M_P^2)_{12}    &=&   (M_P^2)_{21}   = -m_{12}^2 + \lambda_{10} v_1 v_3 + \lambda_{11} v_2 v_3 - \frac{\lambda_{12} v_3^2 }{2} \,,\\
	(M_P^2)_{13}    &=&   (M_P^2)_{31}   = -m_{13}^2 + \lambda_{10} v_1 v_2 + \lambda_{12} v_2 v_3 - \frac{ \lambda_{11} v_2^2}{2} \,,\\      
	(M_P^2)_{23}   &=&   (M_P^2)_{32}  =  -m_{23}^2 + \lambda_{11} v_1 v_2 + \lambda_{12} v_1 v_3 - \frac{ \lambda_{10} v_1^2}{2} \,.        
	\end{eqnarray}
\end{subequations}
For Eq.~\eqref{e:mceqmp} to hold for any arbitrary values of the VEVs, we should have 
\begin{eqnarray}
	\label{e:z3_conditions}
	\lambda_7=\lambda_8=\lambda_9=\lambda_{10}=\lambda_{11}=\lambda_{12}=0\, ,
\end{eqnarray}
which should be read as the conditions for custodial invariance in a $Z_3$ symmetric 3HDM potential.

\subsubsection{The case with a $Z_2\times Z_2'$ symmetry}
\label{ss:Z2xZ2}
The scalar potential in this case can be written as~\cite{Keus:2013hya}
\begin{eqnarray}
	\label{e:z2xz2_pot_v2}
\!\!\!	\!\!\!	\!\!\!	 V_{\rm{Z_2 \times Z_2}}&=&
	m_{11}^2 \phi_1^\dagger \phi_1 + m_{22}^2 \phi_2^\dagger \phi_2 + m_{33}^2 \phi_3^\dagger \phi_3 \nonumber \\
	&& - m_{12}^2 (\phi_1^\dagger \phi_2 + \phi_2^\dagger \phi_1) - m_{13}^2 (\phi_1^\dagger \phi_3 + \phi_3^\dagger \phi_1) -  m_{23}^2 (\phi_2^\dagger \phi_3 + \phi_3^\dagger \phi_2) \nonumber\\
	&&  + \lambda_1(\phi_1^\dagger \phi_1)^2 +\lambda_2(\phi_2^\dagger \phi_2)^2 + \lambda_3(\phi_3^\dagger \phi_3)^2+  \lambda_4(\phi_1^\dagger \phi_1)(\phi_2^\dagger \phi_2)  
	 + \lambda_5(\phi_1^\dagger \phi_1)(\phi_3^\dagger \phi_3) \nonumber \\
	 && + \lambda_6(\phi_2^\dagger \phi_2)(\phi_3^\dagger \phi_3) + \lambda_7(\phi_1^\dagger \phi_2)(\phi_2^\dagger \phi_1) 
	 + \lambda_8(\phi_1^\dagger \phi_3)(\phi_3^\dagger \phi_1) +\lambda_9(\phi_2^\dagger \phi_3)(\phi_3^\dagger \phi_2) \nonumber\\ 
	&&  + \lambda_{10}\left\{(\phi_1^\dagger \phi_2)^2+(\phi_2^\dagger \phi_1)^2 \right\} + \lambda_{11}\left\{(\phi_1^\dagger \phi_3)^2+(\phi_3^\dagger \phi_1)^2 \right\} 
	 + \lambda_{12}\left\{ (\phi_2^\dagger \phi_3)^2+(\phi_3^\dagger \phi_2)^2 \right\},
\end{eqnarray}
where, again, we have allowed terms that softly-break the symmetry.
The elements of the charged-scalar mass matrix are given below:
	\begin{subequations}
		\begin{eqnarray}
		\label{e:Mcsqz2}
		(M_C^2)_{11} &=&	\frac{m_{12}^2 v_2}{v_1} + \frac{m_{13}^2 v_3}{v_1} -\lambda_{10}v_2^2 -\frac{ \lambda_7 v_2^2}{2} - \lambda_{11} v_3^2 - \frac{\lambda_8 v_3^2}{2} \,,\\
		(M_C^2)_{22} &=&  \frac{m_{12}^2 v_1}{v_2} +\frac{m_{23}^2 v_3}{v_2} - \lambda_{10} v_1^2 -\frac{\lambda_7 v_1^2}{2}   -\lambda_{12} v_3^2 - \frac{\lambda_9 v_3^2}{2}\,,\\
		(M_C^2)_{33} &=& \frac{m_{13}^2 v_1}{v_3} + \frac{m_{23}^2 v_2}{v_3} - \lambda_{11} v_1^2 - \frac{\lambda_8 v_1^2}{2}  - \lambda_{12} v_2^2 - \frac{\lambda_9 v_2^2}{2} \,,\\
		(M_C^2)_{12} &=& (M_C^2)_{21} =  -m_{12}^2 + \lambda_{10} v_1v_2 + \frac{1}{2}\lambda_7 v_1v_2 \,,\\
		(M_C^2)_{13} &=& (M_C^2)_{31} =  -m_{13}^2 + \lambda_{11} v_1v_3 + \frac{1}{2}\lambda_8 v_1v_3\,,\\
		(M_C^2)_{23} &=& (M_C^2)_{32} =  -m_{23}^2 + \lambda_{12} v_2v_3 + \frac{1}{2}\lambda_9v_2v_3\,.
		\end{eqnarray}
	\end{subequations}
For the case of the pseudoscalar mass matrix elements, we find
\begin{subequations}
		\begin{eqnarray}
		\label{e:Mpsqz2}
		(M_P^2)_{11} &=& \frac{m_{12}^2 v_2}{v_1} +\frac{m_{13}^2 v_3}{v_1} -2 \lambda_{10}v_2^2 -2 \lambda_{11}v_3^2 \,,	\\
		(M_P^2)_{22} &=& \frac{m_{12}^2 v_1}{v_2} + \frac{m_{23}^3 v_3}{v_2} -2 \lambda_{10} v_1^2  -2 \lambda_{12} v_3^2 \,,\\
		(M_P^2)_{33} &=& \frac{m_{13}^2 v_1}{v_3} + \frac{m_{23}^2 v_2}{v_3} -2 \lambda_{11} v_1^2 -2 \lambda_{12} v_2^2 \,,\\
		(M_P^2)_{12} &=& (M_P^2)_{21}   = -m_{12}^2 + 2 \lambda_{10} v_1 v_2   \,,\\
		(M_P^2)_{13} &=& (M_P^2)_{31}   = -m_{13}^2 + 2 \lambda_{11} v_1 v_3 \,,\\
		(M_P^2)_{23} &=& (M_P^2)_{32}  =  -m_{23}^2 + 2 \lambda_{12} v_2 v_3\,.
		\end{eqnarray}
	\end{subequations}
Following the reasoning presented for the $Z_3$ case, the conditions for custodial invariance can be found using Eq.~\eqref{e:mceqmp}, which read
\begin{eqnarray}
	\label{e:z2z2_conditions}
	\lambda_7=2\lambda_{10}, \,\, \lambda_8=2\lambda_{11}, \, \, \lambda_9=2\lambda_{12}.
\end{eqnarray}
%
%%%%%%%%%%%%%%%%%%%%%%%%%%%%%%%%%%%%%%%%%%%%%%%%%%%%%%%%%%%%%%
\subsection{Wrong-sign Yukawas in democratic 3HDMs}
\label{s:WS}
Now we turn our attention to the Yukawa sector phenomenology that follows from Eq.~\eqref{e:d3hdmlyuk}.  
We will continue to assume that no additional sources for CP violation arise from the scalar sector.  
To begin with, we parametrize the VEVs of the three doublets as follows
\begin{eqnarray}
v_1 = v \cos\beta_1 \cos\beta_2 , \qquad v_2 = v \cos\beta_1\sin\beta_2, \qquad v_3= v \sin\beta_1,
\end{eqnarray}
which, by design, satisfies the relation
\begin{equation}
v_1^2+v_2^2+v_3^2 = v^2, 
\end{equation}
with $v=246$ GeV being the total electroweak VEV.  
The range of values of $\beta_1$ and $\beta_2$ allowed from the perturbativity of the fermionic Yukawa couplings can be found in refs.~\cite{Boto:2021qgu,Chakraborti:2021bpy}.  

The current LHC Higgs data usually serves as a motivation to stay close to the so-called alignment limit~\cite{Das:2019yad}.  
However, as explained in the introduction, here we are after a relatively less-explored possibility where the sign of the down-type Yukawa couplings is opposite to what has been predicted by the SM.  
To prepare ourselves for what comes next, we define the Higgs coupling modifiers as follows~\cite{LHCHiggsCrossSectionWorkingGroup:2012nn, LHCHiggsCrossSectionWorkingGroup:2013rie}
\begin{eqnarray}
\kappa_x = \frac{g_{hxx}}{g_{hxx}^{\rm SM}}\, \, ,
\end{eqnarray}
where the field $h$, in the context of nHDMs, denotes the lightest CP-even scalar, and `$x$' can represent the massive vector bosons or fermions.  

To illustrate the details of the wrong-sign limit, we briefly revisit the example of a type-II 2HDM where the coupling modifiers have the expression given in Table~\ref{tab:kappas}.\footnote{We note here that for the 2HDM case we are using the standard convention for $\alpha$, such that the alignment limit is given by $\cos\left(\beta-\alpha\right)=0$.  
However, for the case of democratic 3HDMs, the angles $\alpha_{1,2}$ are defined in a way such that the alignment conditions read $\sin\left(\alpha_i-\beta_i\right)=0$, with $i=1,2$~\cite{Das:2019yad}.  }  
These coupling modifiers can be conveniently rewritten as follows
\begin{subequations}
\label{eq:kappas2HDM}
\begin{eqnarray}
&&\kappa_V^{\rm II} = \sin\left(\beta-\alpha\right), \\
&&\kappa_u^{\rm II} = \sin\left(\beta-\alpha\right) + \cot\beta \cos\left(\beta-\alpha\right), \\
&&\kappa_d^{\rm II} = \kappa_\ell^{\rm II} = \sin\left(\beta-\alpha\right) - \tan\beta \cos\left(\beta-\alpha\right).
\end{eqnarray}
\end{subequations}
\begin{table}[h]
\centering
\begin{tabular}{c| c c c c}
\hline \hline
\\[-4mm]
Model & $\kappa_V$ & $\kappa_u$ & $\kappa_d$ & $\kappa_\ell$ \\[1mm]
\hline
\\[-4mm]
type-II 2HDM & $\sin\left(\alpha-\beta\right) $  & $\dfrac{\cos\alpha}{\sin{\beta}} $  & $-\dfrac{\sin\alpha}{\cos{\beta}} $  & $-\dfrac{\sin\alpha}{\cos{\beta}} $  \\[3mm]

democratic 3HDMs  & $\displaystyle{\cos\alpha_2\cos\beta_2 \cos\left(\alpha_1-\beta_1\right) \atop + \sin\alpha_2 \sin\beta_2 }$ & $ \dfrac{\sin\alpha_2}{\sin\beta_2} $ & $ \dfrac{\sin\alpha_1}{\sin\beta_1} \dfrac{\cos\alpha_2}{\cos\beta_2} $ & $ \dfrac{\cos\alpha_1}{\cos\beta_1} \dfrac{\cos\alpha_2}{\cos\beta_2} $ \\[3mm]
\hline
\end{tabular}
\caption{\label{tab:kappas} The coupling modifiers for the type-II 2HDM and democratic 3HDMs. In the 2HDM case, $\tan\beta=v_2/v_1$ and $\alpha$ is a suitably defined rotation angle in the CP-even scalar sector~\cite{Branco:2011iw}. Similarly, in the case of 3HDMs, $\alpha_1$ and $\alpha_2$ are two suitably defined rotation angles in the CP-even scalar sector~\cite{Das:2019yad}.  }
\end{table}
%

%
%\begin{wrapfigure}{r}{0.5\textwidth}
\begin{figure}
	\centering
	\includegraphics[width=0.45\textwidth]{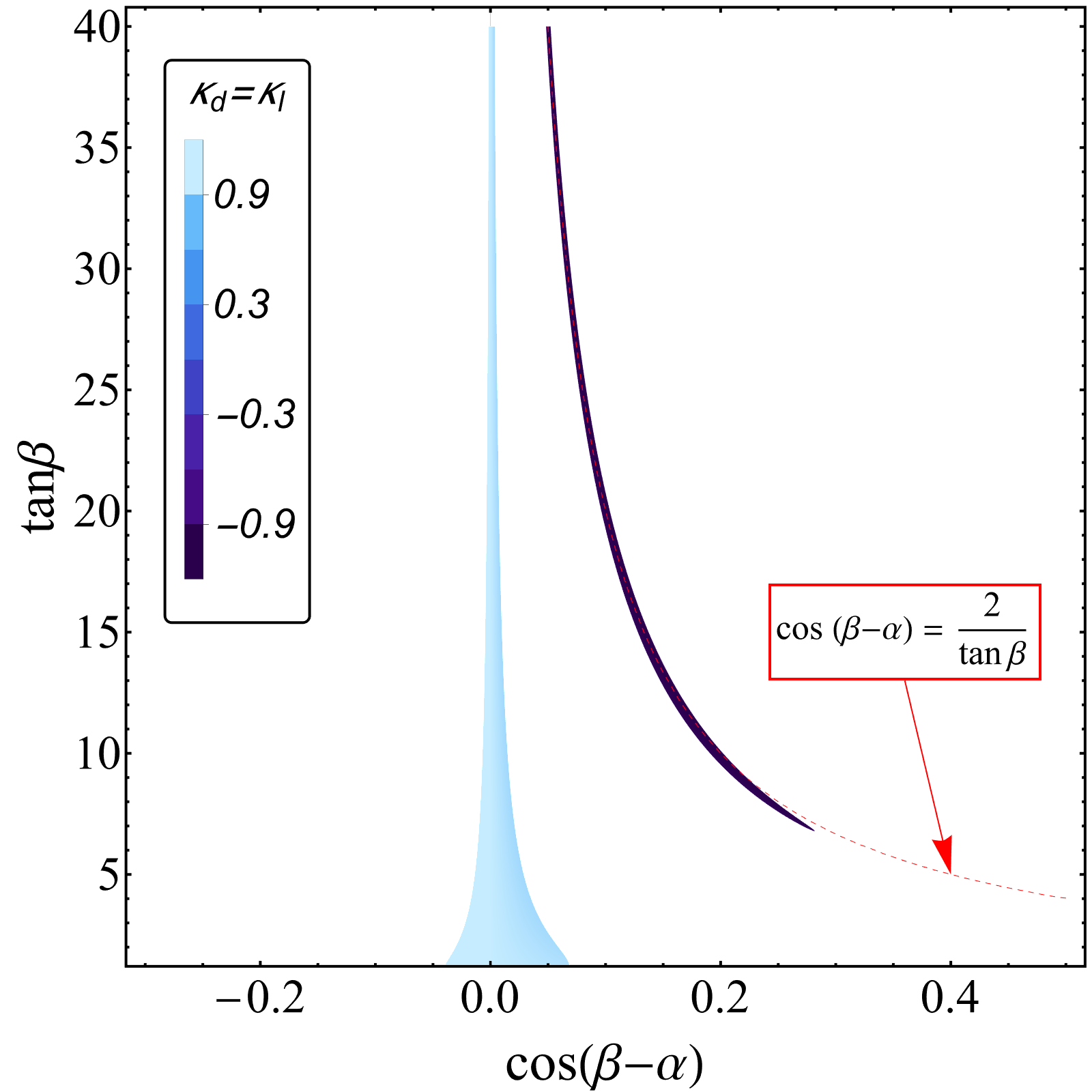}
	\caption{\label{fig:WS2HDM}
		Allowed region at 95\% CL from the current data on Higgs signal strengths in the type-II 2HDM.  
		It should be noted that when considering the $h\to\gamma\gamma$ decay, the charged-Higgs contribution has been neglected with the understanding that it can be safely decoupled in the presence of the soft-breaking parameter in the scalar potential~\cite{Bhattacharyya:2014oka, Carrolo:2021euy,Faro:2020qyp}.  
		For illustration, the line corresponding to \mbox{$\cos(\beta-\alpha)=2/\!\tan\beta$} has also been plotted in the same graph, which reinforces our intuitions from Eq.~\eqref{eq:wsdef2HDM}.  }
\end{figure}
%\end{wrapfigure}
%

%
Now let us consider the limit 
\begin{equation}
\label{eq:wsdef2HDM}
\cos\left(\beta-\alpha\right) = \dfrac{r}{\tan\beta},
\end{equation} 
where $r$ is a real number and $\tan\beta \gg \lvert r \rvert$.  
In such a scenario, Eq.~\eqref{eq:kappas2HDM} can be approximated as
\begin{eqnarray}
\label{e:kappas2approx}
\kappa_V^{\rm II} \approx  1, \qquad
\kappa_u^{\rm II} \approx 1, \qquad
\kappa_{d,\ell}^{\rm II} \approx 1-r.
\end{eqnarray}

The wrong-sign limit, in particular, arises for $r=2$, in which case Eq.~\eqref{e:kappas2approx} takes the following form
\begin{eqnarray}
\label{e:WS2hdm}
\kappa_V^{\rm II} \approx  1, \qquad
\kappa_u^{\rm II} \approx 1, \qquad
\kappa_{d,\ell}^{\rm II} \approx -1.
\end{eqnarray}
Such a possibility is allowed because the current LHC Higgs data is not sensitive enough to probe the sign of the bottom-quark Yukawa coupling in the loop-induced vertices such as $hgg$ and $h\gamma\gamma$.  
To demonstrate this explicitly, we use the current Higgs data~\cite{ATLAS:2021vrm}, and display the $2\sigma$-allowed region in the $\cos\left(\beta-\alpha\right)$ vs $\tan\beta$ plane in Fig.~\ref{fig:WS2HDM}.  
The dark-blue region corresponds to the wrong-sign limit in the type-II 2HDM.\footnote{In a recent 2HDM fit~\cite{Atkinson:2021eox}, it was claimed that the wrong-sign limit is disfavoured by the current Higgs data at $2\sigma$, and only allowed within $3\sigma$.  
However, we have used a more updated dataset and our result for 2HDM agrees with the most updated fit from ATLAS~\cite{ATLAS:2021vrm} (in Fig.~20b, we can see the wrong-sign limit is still allowed).}  

\begin{figure}[h]
\centering
\includegraphics[width=0.45\textwidth]{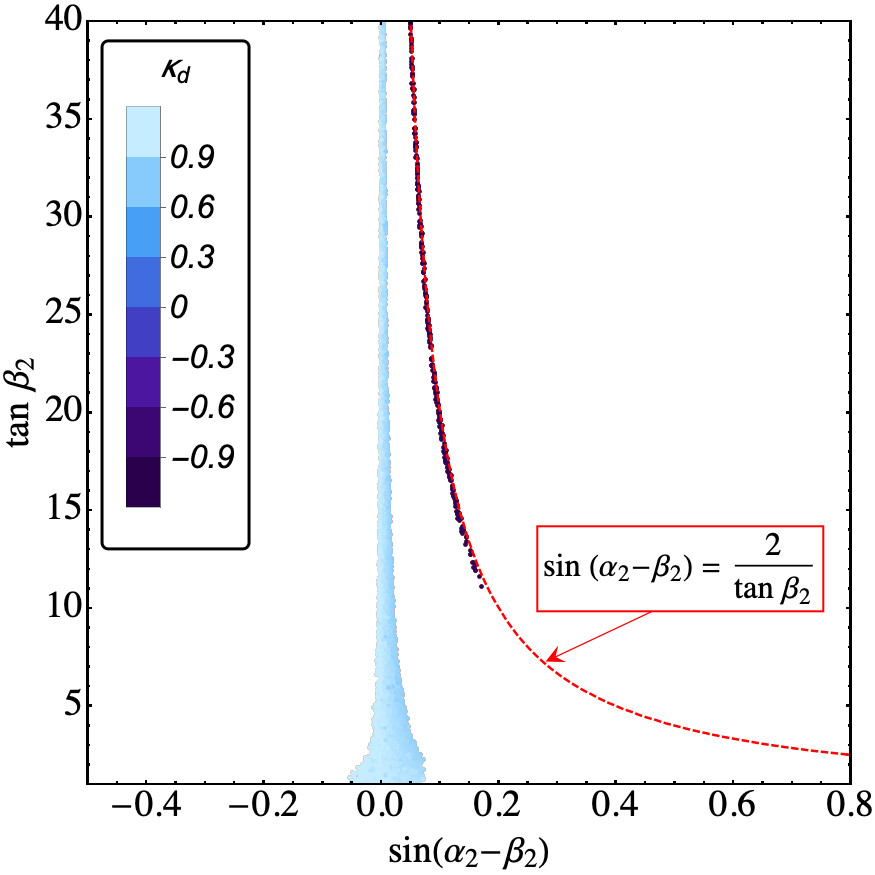} 
~~~~~~~~~~~
\includegraphics[width=0.45\textwidth]{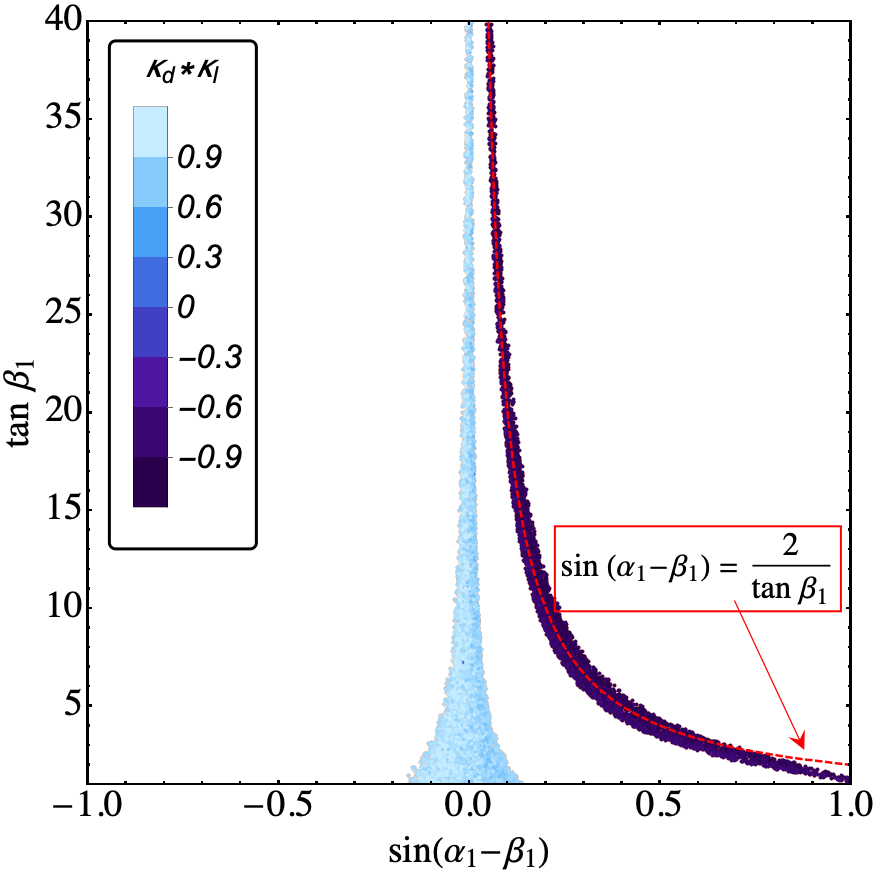}
\caption{\label{fig:WS3HDM-1}
Allowed region at 95\% CL from the current data on Higgs signal strengths in democratic 3HDM.  
%It should be noted that when considering the $h\to\gamma\gamma$ decay, the charged-Higgs contribution has been neglected 
As before, the charged-Higgs contribution to  $h\to\gamma\gamma$ decay is neglected with the understanding that it can be safely decoupled in the presence of the soft-breaking parameter in the scalar potential~\cite{Bhattacharyya:2014oka, Carrolo:2021euy,Faro:2020qyp}.   
The contour corresponding to Eqs.~\eqref{e:WS2} and \eqref{e:WS1}, for $r_1=r_2=2$ are also displayed for easy comparison.  }
\end{figure}

\begin{figure}[h]
\centering
\includegraphics[width=0.45\textwidth]{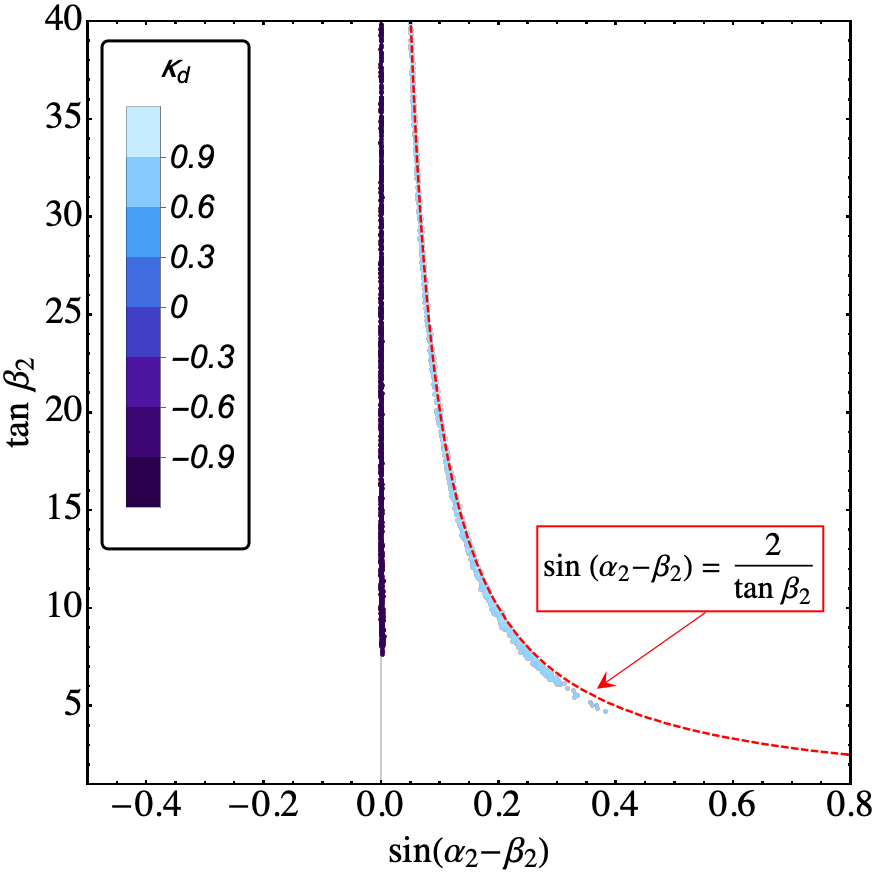}
~~~~~~~~~~~
\includegraphics[width=0.46\textwidth]{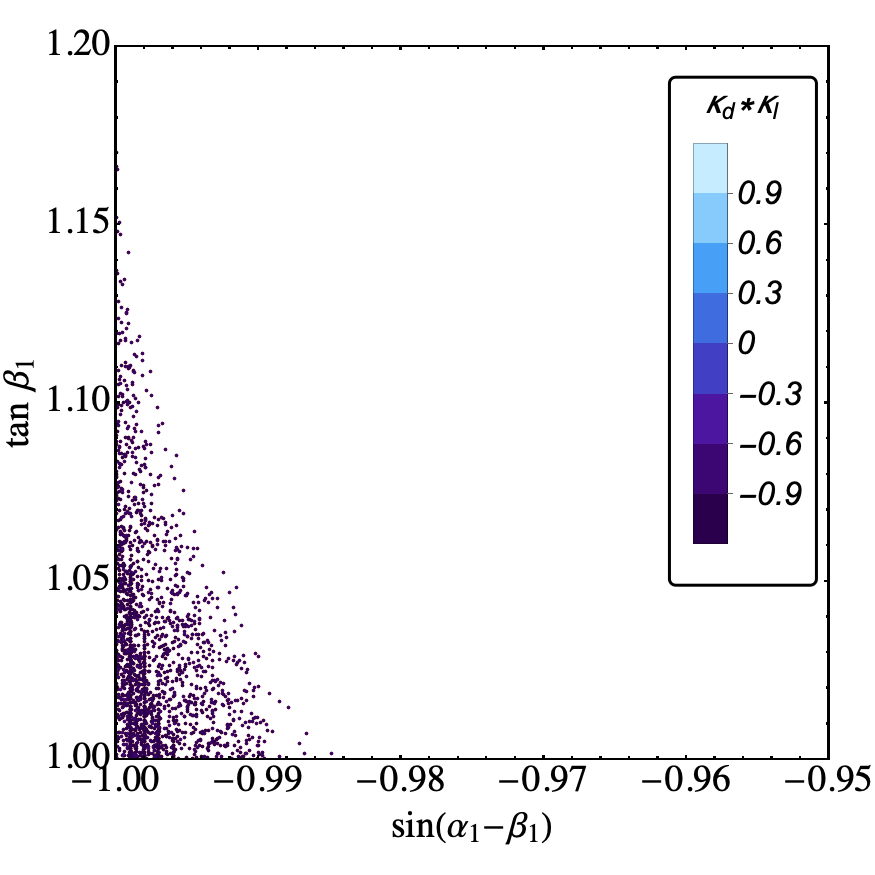}
\caption{\label{fig:WS3HDM-2}
Allowed region at 95\% CL from the current data on Higgs signal strengths  for \mbox{$\sin(\alpha_1 - \beta_1) \approx -1$} is displayed separately  in this plot.
All the points shown in the left panel in the $\sin(\alpha_2 - \beta_2)$ vs.\ $\tan\beta_2$ plane are sampled from the $\sin(\alpha_1 - \beta_1) \approx -1$  region as displayed in the right panel.
%It should be noted that when considering the $h\to\gamma\gamma$ decay, the charged-Higgs contribution has been neglected with the understanding that it can be safely decoupled in the presence of the soft-breaking parameter in the scalar potential~\cite{Bhattacharyya:2014oka, Carrolo:2021euy,Faro:2020qyp}.   
The contour corresponding to Eq.~\eqref{e:WS2} for $r_2=2$ is displayed for easy comparison.  }
\end{figure}

Now, we will demonstrate that such wrong-sign scenarios are also entertained in democratic 3HDMs with much greater flexibility in terms of the number of free parameters.  
To illustrate this, we again purposefully rewrite the Higgs coupling modifiers in Table~\ref{tab:kappas} for democratic 3HDMs as follows  
\begin{subequations}
\label{eq:3hdmkappas}
\begin{eqnarray}
\kappa_V &=& \dfrac{\cos\left(\alpha_1-\beta_1\right)}{1+\tan^2\beta_
2} \bigg( \cos\left( \alpha_2 -\beta_2\right) - \sin\left( \alpha_2 -\beta_2\right) \tan\beta_2 \bigg) \nonumber \\
&& \quad + \quad  \dfrac{\tan^2\beta_2}{1+\tan^2\beta_2} \bigg( \cos\left(\alpha_2-\beta_2\right) + \sin\left(\alpha_2-\beta_2\right) \cot\beta_2\bigg), \\
\kappa_u &=&  \cos\left( \alpha_2 -\beta_2\right) +\sin\left( \alpha_2-\beta_2\right)\cot\beta_2, \\
\kappa_d &=& \bigg( \cos\left( \alpha_1-\beta_1\right) + \sin\left( \alpha_1-\beta_1\right) \cot\beta_1 \bigg) \bigg(\cos\left( \alpha_2-\beta_2\right) - \tan\beta_2 \sin\left(\alpha_2-\beta_2\right) \bigg), \\
\kappa_\ell &=& \bigg( \cos\left( \alpha_1-\beta_1\right) - \sin\left( \alpha_1-\beta_1\right) \tan\beta_1 \bigg) \bigg(\cos\left( \alpha_2-\beta_2\right) - \tan\beta_2 \sin\left(\alpha_2-\beta_2\right) \bigg).
\end{eqnarray}
\end{subequations}
In a similar way to the 2HDM scenario, we focus our attention to the limit 
\begin{equation}
\label{e:WS2}
\sin\left(\alpha_2-\beta_2\right) = \dfrac{r_2}{\tan\beta_2},  
\end{equation}
where $r_2$ is a real number, and $\tan\beta_2 \gg \lvert r_2 \rvert$.  
In this limit, $\kappa_V\approx\kappa_u\approx1$, but $\kappa_d$ and $\kappa_\ell$ take the following form
\begin{subequations}
\label{e:kappas3hdmapprox}
\begin{eqnarray}
\kappa_d &=& (1-r_2) \bigg( \cos\left( \alpha_1 -\beta_1\right) + \sin\left( \alpha_1 - \beta_1\right) \cot\beta_1 \bigg) = (1-r_2) \frac{\sin\alpha_1}{\sin\beta_1}, \\
\kappa_\ell &=& (1-r_2) \bigg( \cos\left( \alpha_1 -\beta_1\right) - \sin\left( \alpha_1 - \beta_1\right) \tan\beta_1 \bigg) = (1-r_2) \frac{\cos\alpha_1}{\cos\beta_1}.
\end{eqnarray}
\end{subequations}
If we further consider the limit
\begin{equation}
\label{e:WS1}
\sin\left(\alpha_1-\beta_1\right) = \frac{r_1}{\tan\beta_1}, 
\end{equation}
where, again, $r_1$ is a real number, and $\tan\beta_1 \gg \lvert r_1\rvert$, then Eq.~\eqref{e:kappas3hdmapprox} can be further simplified to 
\begin{subequations}
\label{e:ws3hdm}
\begin{eqnarray}
\kappa_d &=& (1-r_2), \\
\kappa_\ell &=& (1-r_2)(1-r_1).
\end{eqnarray}
\end{subequations}
The limits that can be obtained for different values of $r_1$ and $r_2$ have been listed in Table~\ref{tab:3HDM wrong sign}, where we can see that all the wrong-sign possibilities that can be obtained from 2HDMs with NFC are encompassed by a democratic 3HDM.  
All these features have been clearly depicted in Figs.~\ref{fig:WS3HDM-1} and~\ref{fig:WS3HDM-2}, where the darker shade correspond to the wrong-sign limit.  
Thus, we can see that the democratic 3HDM gives more leeway for the wrong-sign limit, when compared to the 2HDM.


\begin{table}[h]
\centering
\begin{tabular}{c | c | c }
\hline\hline
& $r_1=0$ & $r_1=2$\\
\hline
$r_2=0$ &  $\begin{matrix} \kappa_d = 1 \quad \kappa_\ell = 1 \\ \text{(alignment limit)} \end{matrix} $ &  $\begin{matrix} \kappa_d \approx 1 \quad \kappa_\ell \approx -1  \\ \text{(wrong-sign limit in the type-X 2HDM)} \end{matrix}$  \\
\hline
$r_2=2$ &  $\begin{matrix} \kappa_d \approx -1 \quad \kappa_\ell \approx -1 \\ \text{(wrong-sign limit in the type-II 2HDM)} \end{matrix}$   &  $\begin{matrix} \kappa_d \approx -1 \quad \kappa_\ell \approx 1 \\ \text{(wrong-sign limit in the type-Y 2HDM)} \end{matrix}$    \\
\hline
\end{tabular}
\caption{\label{tab:3HDM wrong sign} Wrong-sign possibilities in democratic 3HDMs.  
It should be noted that $\kappa_u \approx \kappa_V \approx 1$ in all the cases.  }
\end{table}

So far, we have obtained the wrong-sign limit in the democratic 3HDM following the 2HDM prescription.  
However, a democratic Yukawa structure can entertain more exotic possibilities.  
As usual, we start by investigating how to impose $\kappa_u \approx 1$.  
One possibility is to set $\tan\beta_2 \gg 1$ together with $\cos\left(\alpha_2 -\beta_2\right) \approx 1$, as was done in Eq.~\eqref{e:WS2}, leading to Eq.~\eqref{e:kappas3hdmapprox}.  
Now, instead of going to the limit of Eq.~\eqref{e:WS1}, one can choose 
\begin{eqnarray}
\sin\left(\alpha_1 - \beta_1\right) \approx \pm 1 \, , \qquad \tan\beta_1 \approx 1\, .
\end{eqnarray}
In this way, using $\cos\left(\alpha_1-\beta_1\right) \approx 0$, we get
\begin{subequations}
\begin{eqnarray}
\kappa_V \approx \kappa_u &\approx& 1 , \\
\kappa_d \approx -\kappa_\ell &\approx& \pm \left(1-r_2\right) ,
\end{eqnarray}
\end{subequations}
where, as before, $r_2\approx 0$ and $r_2\approx 2$ can give us two different
possibilities.
As such, we see that it is possible to achieve a wrong-sign limit in the democratic 3HDMs without the requirement of large $\tan\beta_1$.  
If we follow the usual path to the wrong-sign limit, we see that $\sin\left(\alpha_1-\beta_1\right) \approx 1$ is allowed in Fig.~\ref{fig:WS3HDM-1}.  
The possibility with $\sin\left(\alpha_1-\beta_1\right) \approx -1$ is separately showcased in Fig.~\ref{fig:WS3HDM-2} for better visibility.  
%It is interesting to note that in this case, where $\delta=1$, the wrong-sign region for $\kappa_d$ is flipped, \textit{i.e.}, $\kappa_d$ approaches the SM value for $r_2=2$, and the wrong-sign limit is achieved for $r_2=0$.  
%

%\textcolor{red}{
%In the $\sin(\alpha_1 -\beta_1)$ vs $\tan\beta_1$ plane, we note that there are allowed points close to $\sin(\alpha_1-\beta_1) \approx \pm 1$.  
%To understand this, let us consider the limit $\kappa_u\approx 1$ but
%$\sin(\alpha_1-\beta_1)\approx \pm 1$, {\it i.e.}, $\cos(\alpha_1-\beta_1)\approx 0$.
%In such a scenario, if we additionally have $\tan\beta_2 \gg 1$ but $\tan\beta_1\approx 1$,
%then \Eqn{eq:3hdmkappas} can be approximated as 
%%
%\begin{subequations}
%\begin{eqnarray}
%&&\kappa_V \approx \kappa_u \approx 1\, , \\
%&&\kappa_d \approx -\kappa_\ell \approx  (\pm 1)(-1)^\delta \, ,
%\end{eqnarray}
%\end{subequations}
%%
%where the $\pm$ sign corresponds to $\sin(\alpha_1-\beta_1) \approx \pm 1$ and $\delta$
%captures the effect of how $\kappa_u\approx 1$ is achieved. If $\kappa_u\approx 1$ is
%obtained via the usual alignment limit then $\delta=0$, whereas if $\kappa_u\approx 1$
%is realized via the wrong-sign limit, then $\delta=1$.
%%
%However, one should remember that the wrong-sign Yukawas in the vicinity of 
%$\sin(\alpha_1-\beta_1) \approx \pm 1$ require $\tan\beta_1\approx 1$ which might
%be detrimental for sub-Tev nonstandard scalars (?).  
%}

At this point it will be quite natural to wonder how such
wrong-sign possibilities can be probed in experiments. An obvious way to sense the
wrong-sign limit will be to measure the Higgs signal strengths that involve $hgg$
and $h\gamma\gamma$ effective vertices with increasing precision to the extent
that the interference terms from the lighter fermions in the loop start to become
relevant. Alternatively, the decay $h\to\Upsilon\gamma$ was suggested as a probe
for the sign of $\kappa_b$\cite{Modak:2016cdm,Batra:2022wsd}. Similarly $h\to \tau^+\tau^-\gamma$~\cite{Galon:2016ngp} may 
serve as a probe for the sign of $\kappa_\tau$.
Additionally, if we know the UV complete model responsible for the wrong-sign Yukawas,
then we can perform a targeted search for the nonstandard particles. For instance,
in this case the wrong-sign limit is arising within an nHDM framework. Thus, one can
look for nonstandard scalars whose phenomenologies in the wrong-sign limit will be
presumably different from the corresponding alignment limit counterparts\cite{Kanemura:2022cth}.

But the crucial point is, even if we stay agnostic about the origin of the wrong-sign
Yukawas, we should still remember that any departure from the SM couplings will introduce
an energy scale beyond which unitarity will be violated\cite{Joglekar:1973hh}. Therefore, the wrong-sign limits
as described in, {\it e.g.}, \Eqn{e:WS2hdm} will inevitably call for NP below the unitarity
violation scale. For the arrangement of couplings appearing in \Eqn{e:WS2hdm},
the earliest onset of unitarity violation will occur in the $b\bar{b}\to W_LW_L$
scattering and the maximum energy cut-off before which the NP must intervene,
will be given by\cite{Bhattacharyya:2012tj},
\begin{eqnarray}
	E_{\rm max} = \frac{2\sqrt{2}\pi}{G_F m_b} \approx 180~{\rm TeV} \,.
\end{eqnarray}

%
%%%%%%%%%%%%%%%%%%%%%%%%%%%%%%%%%%%%%%%%%%%%%%%%%%%%%%%%%
%%%%%%%%%% Conclusions   %%%%%%%%%%%%%%%%%%%%%
%%%%%%%%%%%%%%%%%%%%%%%%%%%%%%%%%%%%%%%%%%%%%%%%%%%%%%%%
\section{Summary}
\label{s:summary}
In this article we have studied two new aspects of democratic 3HDMs, namely, the
impact of custodial symmetry and the wrong-sign Yukawa couplings.
As such, our goal is to provide the ingredients for constructing democratic 3HDMs 
which is safeguarded against the $T$-parameter constraints, while showcasing 
the interesting Yukawa structure allowed by the Higgs data.
 The custodial
limit serves as a systematic guideline for alleviating the stringent constraints
arising from the electroweak \mbox{$T$-parameter}. We have followed an alternative approach
to find the general condition for the custodial symmetry to be prevalent in
scalar sector of an nHDM. We used these results to extract the model specific
conditions for democratic 3HDMs which usually comes in two different avatars --
one with a $Z_3$ symmetry and the other with a $Z_2\times Z_2'$ symmetry.
We then turn our attention to the Yukawa sector of democratic 3HDMs and showed
that the democratic 3HDMs also accommodate the possibility of wrong-sign limit
where the signs of the down-type Yukawa couplings are opposite to the corresponding
SM predictions. We find that a democratic 3HDM covers all the wrong-sign scenarios
that can possibly arise from a 2HDM framework with NFC. In the recent fits of the
Higgs couplings\cite{ATLAS:2021vrm,ATLAS:2019nkf, CMS:2020eni} in the kappa formalism\cite{LHCHiggsCrossSectionWorkingGroup:2012nn, LHCHiggsCrossSectionWorkingGroup:2013rie}, the results are
often reported with an implicit assumption about the signs of the kappas. Our
discussion on the wrong-sign limit highlights the importance of presenting the
fit results without any inherent assumptions about the signs of the kappas because,
otherwise we can miss potentially interesting and unconventional limits brought
in by many different BSM scenarios. To emphasize the last point, we have also argued
how the wrong-sign limit inevitably leads to an upper limit on the energy scale
for the onset of NP.

%Starting from the scalar kinetic Langrangian, we were able to arrive at a condition that must be satisfied in a CS invariant nHDM without considering the specific form of the scalar potential. The aforementioned condition is the equality of charged scalar and pseudoscalar mass matrices. This result is independent of the basis we choose for the scalars. Since this result is in terms of the masses and mixings of the scalars and is readily applicable to pratical problems. To demonstrate the usefulness of this result, we apply it to calculate that the $T$ parameter satisfies $T=0$ in CS invariant nHDMs.
%%%%%%%%%%%%%%%%%%%%%%%%%%%%%%%%%%%%%%%%%%%%%%%%%%%%%%%%%
%%%%%%%%%% Acknowledgments   %%%%%%%%%%%%%%%%%%%%%
%%%%%%%%%%%%%%%%%%%%%%%%%%%%%%%%%%%%%%%%%%%%%%%%%%%%%%%%
\section*{Acknowledgments}
We thank the anonymous referee for pointing out the missing term in Eq.~\eqref{e:quartic}, which is now included in Eq.~\eqref{e:Nishi}.
DD and AS thank the Science and Engineering Research Board, India for financial support through grant no.  SRG/2020/000006. 
DD and PBP also acknowledge the support from grant no. CRG/2022/000565.
IS acknowledges support from DST-INSPIRE, India, under grant no. IFA21-PH272.
ML acknowledges funding from Funda\c{c}\~{a}o para a Ci\^{e}ncia e a 
Tecnologia (FCT) through Grant No.PD/BD/150488/2019, in the 
framework of the Doctoral Programme IDPASC-PT, and was supported 
in part by FCT projects CFTP-FCT Unit 777 (UID/FIS/00777/2019), 
CERN/FIS-PAR/0008/2019 and CERN/FIS-PAR/0002/2021 which are 
partially funded through POCTI (FEDER), COMPETE, QREN and EU.
AMP acknowledges support from the Govt. of India through the UGC-JRF fellowship.
DD and IS also thank ICTS, Bengaluru for the warm hospitality
while the final stages of this work were being completed.
\appendix 
%%%%%%%%%%%%%%%%%%%%%%%%%%%%%%%%%%%%%%%%%%%%%%%%%%%%%%%%%
%%%%%%%%%% SU(2) Group Theory   %%%%%%%%%%%%%%%%%%%%%
%%%%%%%%%%%%%%%%%%%%%%%%%%%%%%%%%%%%%%%%%%%%%%%%%%%%%%%%
\section{Brief note on $SU(2)$ triplets}
\label{s:SU(2)GT}
A real triplet of $SU(2)$ in the cartesian basis is expressed as follows:
\begin{eqnarray}
\label{e:tripdefsu2}
{\bf A}_{\rm Car}=\begin{pmatrix}
A_1\\A_2\\A_3
\end{pmatrix}
\end{eqnarray}
The generators of $SU(2)$ in this basis are given by
\begin{eqnarray}
\label{e:gendef}
T_1=\begin{pmatrix}
0&0&0\\0&0&-i\\0&i&0
\end{pmatrix}\,,\quad
T_2=\begin{pmatrix}
0&0&i\\0&0&0\\-i&0&0
\end{pmatrix}\,,\quad
T_3=\begin{pmatrix}
0&-i&0\\i&0&0\\0&0&0
\end{pmatrix}\,.
\end{eqnarray}
which make the transformation real. Now we want to migrate to a basis where $T_3$ is diagonal.
We will call this the spherical basis and the $SU(2)$ triplet in this basis will be denoted
by ${\bf A}_{\rm Sph}$.
We note that the unitary matrix
\begin{eqnarray}
\label{e:gendef1}
\mathcal{U}=\frac{1}{\sqrt{2}}\begin{pmatrix}
-1&i&0\\0&0&\sqrt{2}\\1&i&0
\end{pmatrix}\,,
\end{eqnarray}
diagonalizes $T_3$ as follows
\begin{eqnarray}
\label{e:gendef2}
\mathcal{U}\cdot T_3 \cdot\mathcal{U}^\dagger=\begin{pmatrix}
1&0&0\\0&0&0\\0&0&-1
\end{pmatrix}=T_3^\prime\,.
\end{eqnarray}
This implies that ${\bf A}_{\rm Sph}$ will be related to ${\bf A}_{\rm Car}$ via the following relation
\begin{eqnarray}
\label{e:arel}
\bf{A}_{\rm{Sph}} = \mathcal{U} \bf{A}_{\rm{Car}} = \begin{pmatrix}
	\frac{1}{\sqrt{2}} (-A_1+iA_2) \\A_3\\\frac{1}{\sqrt{2}} (A_1 + iA_2)
	\end{pmatrix}\,.
\end{eqnarray}
where we have used Eq.~\eqref{e:tripdefsu2}.  
Now let us define 
\begin{eqnarray}
\label{e:apm}
A_{\pm}= \frac{1}{\sqrt{2}}(A_1\mp i A_2)\,,
\end{eqnarray}
where $A_+$ and $A_-$ are implicitly understood to be the complex conjugates of each other.
In terms of these we can write the $SU(2)$ triplet in the spherical basis as follows 
\begin{eqnarray}
\label{e:tripletdefsu2}
A_{\rm{Sph} }=\begin{pmatrix}
-A_+\\A_3\\A_-
\end{pmatrix}\,.
\end{eqnarray}
Thus, the $SU(2)$ invariant combination of two triplets, in these two bases, will be given by
\begin{subequations}
	\label{e:inv}
	\begin{eqnarray}
	\textbf{A}\cdot \textbf{B}&=& A_1B_1 + A_2B_2 +A_3B_3\\
	&=& A_+B_- + A_-B_+ + A_3B_3\,.
	\end{eqnarray}
\end{subequations}
In a similar manner, the $SU(2)$ invariant combination of three triplets is expressed as
\begin{subequations}
	\label{e:inv3}
	\begin{eqnarray}
	(\textbf{A}\times \textbf{B})\cdot \textbf{C} &=& (A_2B_3-B_2A_3)C_1+(A_3B_1-B_3A_1)C_2+(A_1B_2-B_1A_2)C_3 \\
	&=& i \left[A_3(B_-C_+ - C_-B_+) +B_3(C_-A_+ -A_-C_+) +C_3(A_-B_+ - B_-A_+)\right] \,.
	\end{eqnarray}
\end{subequations}

%%%%%%%%%%%%%%%%%%%%%%%%%%%%%%%%%%%%%%%%%%%%%%%%%%%%%%%%%
%%%%%%%%%% Comparing with Previous Results   %%%%%%%%%%%%%%%%%%%%%
%%%%%%%%%%%%%%%%%%%%%%%%%%%%%%%%%%%%%%%%%%%%%%%%%%%%%%%%
\section{Custodially invariant scalar potential}
\label{s:compare}
In this Appendix, we try to enumerate the terms in the scalar
potential of a CS-invariant nHDM.  Since we have doublets only, the
renormalizable scalar potential can contain only quadratic and quartic
terms.

In $n$ doublets, there are $4n$ real fields.  After the symmetry
breaking, there will be $n$ triplets of the CS, including one that
contains the unphysical Goldstone modes.  In addition, there will be
$n$ singlets. 
The real parts of the neutral components of $\phi_k$ will be CS
singlets.  It is then easy to see that 
\begin{subequations}
	\label{quad}
	\begin{eqnarray}
	\phi_k^\dagger \phi_k 
	&=& \frac12  {\bf T}_k \cdot {\bf T}_k +
	\mbox{CS singlets}, \\
	\phi_j^\dagger \phi_k + \phi_k^\dagger \phi_j &=& 
	 {\bf T}_j \cdot {\bf T}_k + \mbox{CS singlets},
	%\label{}
	\end{eqnarray}
\end{subequations}
with $j\neq k$.  These are the quadratic forms which are CS
invariant\cite{Pomarol:1993mu,Olaussen:2010aq,Solberg:2018aav}.  The total number of terms of the first kind is $n$, and of
the second kind is $\frac12 n(n-1)$, making a total of $\frac12 n(n+1)$,
which is also exactly the number of different quadratic terms of the form ${\bf T}_j
\cdot {\bf T}_k$ that we can get, with unrestricted $j$ and $k$.  In
fact, if we insist on only real parameters in the scalar potential,
there is no additional restriction arising from the CS: the terms shown in 
\Eqn{quad} are the only ones that are Hermitian and gauge invariant.

A large subset of the quartic CS invariants can be constructed as combinations of the quadratics. We
can enumerate these kinds of terms as follows: 
\begin{subequations}
	\label{e:quartic}
	\begin{eqnarray}
	(\phi_i^\dagger \phi_i)^2 & : \quad : & \mbox{$n$ terms}, \\
	(\phi_i^\dagger \phi_i)(\phi_j^\dagger \phi_j) & : (i\neq j) : &
	\mbox{$N$ terms}, \\ 
	(\phi_i^\dagger \phi_j + \phi_j^\dagger \phi_i)^2
	& : (i\neq j) : & \mbox{$N$ terms}, \\
	(\phi_i^\dagger \phi_j + \phi_j^\dagger \phi_i)(\phi_k^\dagger
	\phi_l + \phi_l^\dagger \phi_k) & : (\{i.j\} \neq \{k,l\})  : &
	\mbox{$\frac12N(N-1)$ terms}, \\ 
	(\phi_i^\dagger \phi_i)(\phi_k^\dagger
	\phi_l + \phi_l^\dagger \phi_k) & : (k\neq l) : & \mbox{$nN$ terms},
	%\label{}
	\end{eqnarray}
\end{subequations}
where, $N=\frac12n(n-1)$. The total number of such terms is $ \frac18n(n+1)(n^2+n+2)$.  
The number of such terms arising from pairs of dot product type combinations of $n$ triplets of CS comes out to be exactly the same.  
For nHDMs with $n\ge4$, as discussed in Ref.~\cite{Nishi:2011gc}, it is possible to obtain a new gauge invariant quantity that is truly independent of the combinations listed in Eq.~\eqref{e:quartic} and corresponding to it we have the following CS invariant:
\begin{eqnarray}\nonumber
	&&\text{Im}(\phi_i^\dagger\phi_j)\text{Im}(\phi_k^\dagger\phi_l)+\text{Im}(\phi_i^\dagger\phi_l)\text{Im}(\phi_j^\dagger\phi_k)+\text{Im}(\phi_i^\dagger\phi_k)\text{Im}(\phi_l^\dagger\phi_j)\\&&=-\frac{1}{4}\left[({\bf T}_i \times {\bf T}_j)\cdot {\bf T}_k h_l -({\bf T}_j \times {\bf T}_k)\cdot {\bf T}_l h_i+({\bf T}_k \times {\bf T}_l)\cdot {\bf T}_i h_j-({\bf T}_l \times {\bf T}_i)\cdot {\bf T}_j h_k \right] 	\label{e:Nishi}
\end{eqnarray}
with $i\ne j\ne k \ne l$. However, the term in \Eqn{e:Nishi} does
	not contribute to the mass matrices and therefore complies with \Eqn{e:mceqmp}.
It should be noted that in the most general gauge invariant
potential, many more quartic terms are possible.  Thus,  
the quartic coefficients, $\lambda_i$, need to be correlated in such a way so that the terms in the quartic part of the scalar potential can be expressed in terms of the $SU(2)_C$ invariant quantities listed in Eqs.~\eqref{e:quartic} and \eqref{e:Nishi}.

To elucidate the implications, let us go back to the example of the 2HDM scalar potential.  
From the general 2HDM potential of Eq.~\eqref{e:2hdmpot1}, we can see that the only terms that are not expressible in terms of the $SU(2)_C$ bilinear invariants of Eq.~\eqref{e:quartic} are the terms proportional to $\lambda_4$ and $\lambda_5$.  
But in the custodial limit of Eq.~\eqref{e:z2_conditions}, these two terms can be combined as 
\begin{eqnarray}
\label{e:gentocs2hdm}
\lambda_4(\phi_1^\dagger \phi_2)(\phi_2^\dagger \phi_1) + \frac{\lambda_5}{2} \left\{(\phi_1^\dagger \phi_2)^2+(\phi_2^\dagger\phi_1)^2\right\}\xrightarrow{\lambda_4=\lambda_5}\frac{\lambda_4}{2}(\phi_1^\dagger \phi_2+\phi_2^\dagger \phi_1)^2
\end{eqnarray}
which, in view of Eq.~\eqref{e:quartic}, is $SU(2)_C$ invariant.
  
The above discussion can easily be extended to the case of nHDMs, especially to the democratic 3HDMs, discussed in section~\ref{s:CS_lim_Dem3HDMs}.  
The conditions obtained using Eq.~\eqref{e:mceqmp} thus rearrange the quartic part of the scalar potential in such a way that it can be expressed as combinations of the terms listed in Eq.~\eqref{e:quartic}.  
\bibliographystyle{JHEP}
\bibliography{Custodial.bib}

\end{document}